\documentclass[10pt,conference]{IEEEtran}

\usepackage[T1]{fontenc}
\usepackage{thm-restate}
\usepackage{amsmath}
\usepackage{amssymb}
\usepackage{mathtools}
\usepackage[utf8]{inputenc}
\usepackage[font=small,labelfont=bf,tableposition=top]{caption}
\usepackage{blindtext}
\usepackage{cuted}
\usepackage{array}

\usepackage{xcolor}
\usepackage{color,balance}
\usepackage{subfiles,paralist}
\usepackage{enumitem}
\usepackage{extarrows}
\usepackage{hyperref}
\usepackage[capitalise]{cleveref}
\crefformat{section}{\S#2#1#3}
\crefformat{subsection}{\S#2#1#3}
\crefformat{subsubsection}{\S#2#1#3}
\crefrangeformat{section}{\S\S#3#1#4 to~#5#2#6}
\crefmultiformat{section}{\S\S#2#1#3}{ and~#2#1#3}{, #2#1#3}{ and~#2#1#3}

\usepackage{subcaption}
\usepackage{graphicx}
\usepackage{adjustbox}
\usepackage[boxruled,linesnumbered, ruled,vlined,noend]{algorithm2e}
\usepackage[noend]{algpseudocode}
\usepackage[most]{tcolorbox}
\usepackage{listings}
\usepackage{cleveref}
\usepackage{placeins}
\usepackage{afterpage}

\usepackage{xpatch}
\xapptocmd\normalsize{%
 \abovedisplayskip=4pt
 \abovedisplayshortskip=0pt plus 4pt
 \belowdisplayskip=4pt
 \belowdisplayshortskip=5pt plus 3pt minus 4pt
}{}{}
\usepackage{diagbox}
\usepackage[normalem]{ulem}
\usepackage{multirow}
\usepackage{colortbl}
\usepackage{wrapfig}
\usepackage{cleveref}
\usepackage[misc,geometry]{ifsym}
\usepackage{cite}
\usepackage{pifont}

\newcommand{\tian}[1]{{\color{black}#1}}

\newcommand{\thomasdir}[1]{{\color{black}#1}}

\newcommand{\player}[1]{{\color{blue}\lambda_{#1}}}

\renewcommand\paragraph[1]{\noindent \textbf{#1.\ }}

\SetAlFnt{\small}
\SetAlCapFnt{\small}
\SetAlCapNameFnt{\small}
\SetAlCapHSkip{0pt}
\IncMargin{-\parindent}

\newtheorem{theorem}{Theorem}[section]
\newtheorem{definition}{Definition}
\newtheorem{proof}{Proof}

\newcommand{\circno}[1]{\textcircled{\raisebox{-0.9pt}{#1}}}

\newif\iffull
\fulltrue 
\iffull
  \newenvironment{equationFC}{\begin{equation*}}{\end{equation*}}
  \newenvironment{itemizeFC}{\begin{itemize}}{\end{itemize}}
  \newenvironment{enumerateFC}{\begin{enumerate}}{\end{enumerate}}
  \newcommand{\FULL}[1]{#1}
  \newcommand{\CONF}[1]{}
\else
\usepackage{enumitem}
  
  \newenvironment{itemizeFC}{\begin{itemize}[leftmargin=*,noitemsep,topsep=0pt]}{\end{itemize}}
  
  \newcommand{\FULL}[1]{}
  \newcommand{\CONF}[1]{#1}
\fi
\usepackage{placeins}
\def\BibTeX{{\rm B\kern-.05em{\sc i\kern-.025em b}\kern-.08em
    T\kern-.1667em\lower.7ex\hbox{E}\kern-.125emX}}

\begin{document}

\title{The Case of FBA as a DEX Processing Model} 
\author{
    \IEEEauthorblockN{Tiantian Gong\IEEEauthorrefmark{1}\IEEEauthorrefmark{2}\thanks{\IEEEauthorrefmark{2}This work is mostly done while the author was a Ph.D. candidate at Purdue University.}, Zeyu Liu\IEEEauthorrefmark{1}, Aniket Kate\IEEEauthorrefmark{3}}
    \IEEEauthorblockA{\IEEEauthorrefmark{1}Yale University, 
    \{tiantian.gong, zeyu.liu\}@yale.edu}
    \IEEEauthorblockA{\IEEEauthorrefmark{3}Purdue University / Supra Research, 
    aniket@purdue.edu}
}

\maketitle

\begin{abstract}
    We investigate the welfare loss of \emph{continuous} and \emph{discrete} order matching models in blockchain-based decentralized exchanges (DEX) that utilize order books to record outstanding orders. 
    \textit{Continuous processing} matches each incoming transaction against the current order book. The \textit{discrete processing} model, i.e., \textit{frequent batch auction} (FBA), executes transactions discretely in batches with a uniform price double auction: Orders are first matched according to \textit{price}, then the exact transaction order if competing orders specify the same price.

    We find that FBA imposes less \textbf{welfare loss} and provides better \textbf{liquidity} than continuous processing in typical scenarios, e.g., when few parties are \textit{privately informed} about asset valuations. 
    Even otherwise, it achieves better social welfare and liquidity provision in the following settings: when price takers and public information reflecting asset value changes arrive sufficiently frequently compared to private information, when the priority fees (for faster transaction inclusion into blockchains) are small, or when the market is more balanced on both buy and sell sides. Our empirical analysis on the BTC-USD and ETH-USD transactions on a DEX named dYdX indicates that FBA can reduce transaction costs by $21\%-37\%$. 
\end{abstract}

\pagestyle{plain}

\section{Introduction}\label{sec:intro}
Blockchain-based decentralized exchanges (DEX) enable users to trade assets without relying on intermediary authorities through blockchain transactions. They are already valued at several billion USD and are expected to grow multi-fold in the coming years. 
DEXes typically take one of the two following forms. (1) In a limit order book-based design (e.g., Penumbra~\cite{sui,penumbra}, Uniswap limit orders~\cite{uniswaporderbook}, CoW Swap~\cite{cow}), each transaction is a limit order (which specifies the direction\footnote{Sell or buy}, price, and quantity of trade), and orders are matched individually. 
A set of outstanding limit orders is called a limit order book. 
(2) An automated market maker design (e.g., Uniswap~\cite{adams2021uniswap}) features automatic algorithmic trading where the settling price ensures certain pre-defined function behaves in a pre-determined way. 

In this paper, we analyze order book-based DEX designs
with a focus on comparing two applicable \textit{transaction processing models}, i.e., \textit{continuous} processing (CLOB, continuous limit order book) where transactions are handled one by one in order or \textit{discrete} frequent batch auction (FBA) where transactions are aggregated at a certain frequency and settled in a double auction (i.e., bids are from both sellers and buyers). 
Similar to the traditional centralized exchanges (CEXs), many order book-based DEX designs follow the CLOB model.
Two main metrics to compare the performance of the two designs are (1) \textit{welfare loss}, which measures the extra transaction costs paid by a common user and is determined by the adverse selection risk (explained below) and the adopted processing model (either CLOB or FBA); and (2) \textit{liquidity provision}, which measures the extent to which an on-demand user's transaction is settled (against the order book)\footnote{The bid-ask spread is usually viewed as a proxy for measuring liquidity.}.
We evaluate the suitability of FBA in the blockchain setting by uncovering the scenarios where FBA provides better welfare and liquidity.

\paragraph{Why consider FBA}
While FBA is less considered,
it can potentially fit the DEX setting better.
In addition to the discussions in the context of centralized exchanges~\cite{farmer2012review,budish2015high}, there are several other intuitive reasons. 
First, inherently, blockchains already treat time as discrete when ordering and assembling transactions into blocks, the same as in FBA. 
Second, when transactions are executed continuously (via CLOB), one can observe and act on the ``future'' if the pending transaction pool is not fully hidden, thus rendering one of CLOB's main advantages moot. 
Similarly, latencies in block generation and message transmission along with priority fees allow even more space for latency arbitrage rents in DEXes, making CLOB even less favorable.
Lastly, a common critique of FBA, the non-execution risks caused by nontransparent pre-trade order book, is alleviated in public blockchains where at least part of the transactions are overt.\footnote{
In centralized exchanges employing FBA~\cite{twse}, quotes are hidden until being matched. For CEX, displaying orders directly could violate certain regulations~\cite{budish2014implementation}, e.g., Rule 610 in Regulation National Market System.
}

\paragraph{Analysis framework}
In the DEX we consider three types of players: (1) \emph{common investors} who submit inelastic trading orders, (2) \emph{informed traders} who submit trades after observing publicly or privately available information about asset valuation jumps, and (3) (many) \emph{arbitrageurs} who can apply market-making (i.e., placing orders to facilitate trades) and front-running strategies. Arbitrageurs are addressed as \emph{liquidity providers} (LPs) when applying the market-making strategy, and as \emph{front-runners} when front-running investors, traders, or other LPs. Our focus is then on how LPs set up their quotes, which decide the amount by which the asking price exceeds the bid price, i.e., the bid-ask spread. 

Putting ourselves in LPs' shoes, 
common investors are price takers and thus the \textit{source of profits}. Informed traders and front-runners cause \textit{losses} to LPs by reacting to information faster, which is also called the \textit{adverse selection risk}~\cite{glosten1985bid,budish2015high}. 
Hence, the arrival rates of common investors and the changes in public or private information about asset valuation affect how LPs determine the bid-ask spread. An LP does not set it to be too high, considering that other arbitrageurs can improve on it (i.e., setting a lower spread) to trade with common investors and earn a profit. The LP does not make it too small either, to absorb the potential loss from adverse selection. 
Here, the \textit{markup} set aside to tolerate this adverse selection risk is the \emph{welfare loss} of our interest. It is considered a ``loss'' of the welfare of common investors and traders because this portion of the spreads does not need to be paid when there is no adverse selection. 
Aside from the welfare loss, we are also interested in \emph{liquidity provision}, which is measured with the bid-ask spread itself. Overall, the spread consists of the markup (i.e., the welfare loss) and the price impact of new incoming orders (because it may reflect public or private information about the underlying asset's price changes).

\subsection{Contributions}
In summary, we make two contributions. First, we analyze the theoretical benefits of employing FBA (frequent batch auction) as the transaction processing model for DEX compared to CLOB (continuous limit order book). Specifically, we uncover typical market conditions where FBA provides better welfare and liquidity. Second, we conduct empirical analysis to straightforwardly compare the transaction costs under the two processing approaches.

\paragraph{Welfare and liquidity provision}
For \underline{welfare loss}, if there is \emph{no} private information in DEXes, then FBA always imposes a \textit{zero} welfare loss while CLOB has a positive welfare loss. 
If a specific DEX design allows privately informed parties, e.g., a board member or a developer capable of dictating protocol updates, adverse selection is present under FBA. In this case, FBA inflicts positive welfare loss. 
In this setting with private information, we find that first, FBA imposes less welfare loss than CLOB if common investors and public information concerning asset value changes arrive sufficiently often as compared to private information. This is because, under FBA, LPs have more time to respond to public information, and less private information reduces adverse selection. 
Second, the smaller the transaction priority fees~\cite{ethfee} are, the higher the markups under CLOB are, as compared to under FBA.~\footnote{Average fee per transaction is \$$1.18$ for Ethereum between September 11-October 10 in 2024, during which the average daily price changes of ETH is \$$49.37$.} 
This is because the profits from front-running increase as costs decline, resulting in the LPs charging higher markups to counter adverse selection. 
Finally, a more balanced market decreases the welfare loss for FBA, making it typically less than the welfare loss under CLOB.
In essence, when sufficiently many transactions are accumulated and matched among themselves, the welfare loss in FBA decreases. 

In terms of \underline{liquidity provision}, FBA has smaller bid-ask spreads (thus better liquidity provision) under similar conditions.

\paragraph{Empirical analysis}
Overall, if the public blockchain that performs the DEX function has much less private information than public information,
FBA provides lower welfare loss and better liquidity. 
More specifically, assuming that the system has little private information,
using real-world transactions acquired from the dYdX exchange~\cite{dydx} from January 2023 to October 2024, we find that empirically there is a $21\%$-$37\%$ increase in transaction costs when the transactions are executed with CLOB compared with using FBA.

\section{Model and definitions}\label{sec:prelim}
In this section, we specify the system and trading system models, and then introduce game theoretic concepts.

\subsection{System model}
A set of $n$ processes called validators run a secure blockchain system that mitigates \textit{transaction order manipulation attacks} with order-fair atomic broadcast (of-ABC)~\cite{cachin2021quick} (\Cref{def:fairabc}). 
Up to $f$ validators are Byzantine and behave arbitrarily. They communicate via reliable authenticated point-to-point channels. More specifically, they are connected with Byzantine fault-tolerant first-in-first-out (FIFO) \textit{consistent broadcast links}~\cite{cachin2021quick} that securely deliver messages. 
The network is \textit{partially synchronous} where the network imposes some known bounded message delay after a global stabilization time (GST).

\begin{definition}[Differentially of-ABC~\cite{cachin2021quick}]
 A secure $\kappa$-differentially of-ABC protocol satisfies:
 \begin{compactenum}
 \item \textbf{Agreement}: If a message $m$ is delivered by some correct process, then $m$ is eventually delivered by every correct process. 
 \item \textbf{Integrity}: No message is delivered more than once.
 \item \textbf{Weak validity}: If all processes are correct and broadcast a finite number of messages, every correct process eventually delivers these broadcast messages. 
 \item \textbf{Total order}: Let $m$ and $m_0$ be two messages, $P_i$ and $P_j$ be correct processes that deliver $m, m_0$. If $P_i$ delivers $m$ before $m_0$, $P_j$ also delivers $m$ before $m_0$.
 \item \textbf{$\kappa$-differential order fairness}: If $b(m; m_0) > b(m_0; m)+2f+\kappa$ (where $b(x; y)$ counts the number of processes that consistent-broadcast $x$ before $y$), then no correct process delivers $m_0$ before $m$. 
 \end{compactenum}\label{def:fairabc}
\end{definition}
Note that we can adopt alternative order manipulation techniques. We only need to re-calibrate the front-running success probability.

\paragraph{Front-running success probability}
Suppose after observing the transaction $m_i$ of a user $i$, arbitrageur $j$ decides to front-run $m_i$ with transaction $m_j$. To compute $j$'s success probability in of-ABC, we assume the distribution of latencies on different communication links is known. 
The links connecting the validators have latency distribution $F_{v}$; the links between common users (i.e., investors and traders) and validators have latency distribution $F_{it}$; the links between arbitrageurs and validators have latency distribution $F_{a}$. 

Let $D$ be the random variable for the latency differences to independently deliver two messages on a communication link between validators. Let $D$ follow distribution $\tilde{F}_{lt}$. For example, if $F_{v}$ is instantiated with normal distribution $N(1,1)$, then $\tilde{F}_{lt}$ follows distribution $N(0,2)$. Let $\tilde{F}^C_{lt}$ be the CDF of $\tilde{F}_{lt}$. 
Assuming the worst case where $j$ sees $m_i$ immediately after it arrives at some validator, $j$ then immediately sends $m_j$ to a validator, and $j$ always wins when neither transaction is broadcast first by more than $(2f+\kappa)$ validators. We denote $j$'s winning probability as $p^{\star}$ and compute it as follows:
\scriptsize
\begin{align*}
 p^{\star} \FULL{&}= 1 - \mathbb{P}[ b(m_i;m_j) > b(m_j;m_i) + 2f + \kappa]\FULL{ \\}
 \FULL{&}= 1 - \sum_{s=2f + \kappa - 1}^{n-1} \binom{n-1}{s} [\tilde{F}^C_{lt}(0)]^s [1-\tilde{F}^C_{lt}(0)]^{n-1-s}
\end{align*}
\normalsize
\FULL{We have $(n-1)$ as the last index instead of $n$ because $m_i$ first arrives at a validator before being observed by arbitrageur $j$.}
If Byzantine validators do not relay messages, we replace $(n-1)$ with $(n-f-1)$. 
In the above example where $F_{v}\sim N(1,1)$ and $\tilde{F}_{lt}\sim N(0,2)$, we have $p^{\star} = 0.75$ for $n=10,f=3,\kappa=1$ and $p^{\star} = 0.95$ for $n=31,f=10,\kappa=1$.

\subsection{Trading system}\label{sec:model}
We adapt the dynamic trading models in Eibelsh{\"a}user and Smetak~\cite{eibelshauser2022frequent} and Budish et al.~\cite{budish2015high} in traditional centralized exchanges to the blockchain-based DEX. 
The frequency of the batch auction is every one or multiple blocks. We denote the length of this period as $I$. 

We consider an asset $X$ with changing fundamental value $V_t$ at time $t$ ($t\in [0, T]$, for some $T>0$). 
We assume an observable signal that equals the fundamental value of the asset and evolves according to a compound Poisson jump process, drawn from a symmetric distribution $F_{jp}$ with arrival rate $\lambda_{jp}$, bounded support, and zero mean. The fundamental value jump can be observed as both \textit{public} and \textit{private} information. 
We capture the absolute value of the jump with the random variable $J$. 

We model three types of risk-neutral\footnote{i.e., obtaining the same amount in expectation generates the same utility.} trading players:
\begin{itemizeFC} 
\item[${\color{gray!50}\blacksquare}$] \textbf{Investors} who arrive stochastically with probability $\player{i}$ and have inelastic demand, with buying and selling being equally likely.
 \item[${\color{gray!50}\blacksquare}$] \textbf{Informed traders} who can observe both the public information arriving stochastically with probability $\player{pb}$ and the private information arriving stochastically with probability $\player{pr}$. Both types of information affect $V_t$ positively or negatively with equal probability.
 \item[${\color{gray!50}\blacksquare}$] $r$ \textbf{arbitrageurs} who can apply market-making or front-running strategy. They can also observe and respond to the public information mentioned above (arriving with probability $\player{pb}$). At time $t$, they intend to sell at a price higher than $V_t$ or buy at a price lower than $V_t$. 
\end{itemizeFC}
The arbitrageurs here can be trading firms and arbitrage bots, among others. As mentioned in \Cref{sec:intro}, a front-running arbitrageur can front-run LPs, investors, and traders. Unlike in the modeling in centralized exchanges, the front-runner now can act after others have already submitted transactions. 

\paragraph{Trading system states}
We capture the history information at time $t$ with $\mathcal{H}_t$, and it is observable by all. The state of the trading system at time $t$ is a tuple $S_t = (P_t, J_t, (\vec{b_t}, \vec{a_t}), \vec{\mathsf{g}})$, where $P_t = \mathbb{E} [V_t|\mathcal{H}_t]$ is the expected value of the fundamental value of the asset given the observed history, $J_t$ is sampled from distribution $J$ (i.e., $J_t \sim J$), $(\vec{b_t}, \vec{a_t})$ are \thomasdir{quotes of} the bid and ask prices, and $\vec{\mathsf{g}}$ records the probabilities that arbitrageurs respond to investors' and traders' orders with front-running at each level of the order book.

\paragraph{Fees}
The exchange collects fees from both sides of each settled trade. 
There are a base fee and a priority fee $\mathsf{F}$ which can promote the order of a transaction during tie-breaking. In the analysis, we disregard the base fee since it is the same for all transactions and is burnt in some systems~\cite{ethfee}.

\subsection{Definitions - solution concepts}
In the trading game, $r$ arbitrageurs are the strategic players and aim to maximize their utilities by playing the best strategy. 
A \textit{strategy} is a probabilistic distribution over possible actions, with all mass condensed at one action for a pure strategy. A \textit{solution concept} then describes a profile or snapshot of all players' strategies with certain desired properties, e.g., Nash equilibrium discourages unilateral deviation. 
We naturally adopt stationary Markov Perfect Equilibrium (MPE)~\cite{maskin2001markov} for CLOB since the parameters of player and information arrival rates in the stochastic game are time-independent.

\begin{definition}[Stationary MPE~\cite{maskin2001markov}]
A Nash equilibrium (NE) is a strategy profile $\vec{s}$ where no player increases utility by unilaterally deviating from $\vec{s}$. 
A subgame perfect equilibrium (SPE) is a strategy profile $\vec{s}$ that forms an NE for any subgame of the original game. 
An MPE is an SPE in which all players play Markov strategies, i.e., strategies that depend only on the current state of the game. A stationary MPE is an MPE where strategies are time-independent.
\end{definition}

We utilize a weaker notion, Order Book Equilibrium (OBE) for FBA since stationary MPE does not always exist for FBA~\cite{budish2019theory}. 
Intuitively, this is because under FBA, when an arbitrageur provides liquidity at the bid-ask spread that equals costs from market-making, other arbitrageurs do not have the incentive to undercut (by improving the current quotes). 
Since others do not undercut, the arbitrageur has the incentive to widen the spread to increase profits, which is a unilateral action to increase the player's utility. But this would result in others undercutting the widened quotes. 
\thomasdir{We first define OBE and discuss why adopting the weaker notion still gives a fair comparison.}

\begin{definition}[OBE~\cite{budish2019theory}]
Given state $S_t$, an OBE at time $t$ is a set of orders submitted by all arbitrageurs such that the following hold: There exist (1) no safe profitable price improvements and (2) no other safe profitable deviations. 
\end{definition}
Here, a profitable price improvement is safe if it remains strictly profitable after other arbitrageurs take profitable responses, e.g., liquidity withdrawals, after the improvement. A profitable deviation is safe if it remains strictly profitable after other arbitrageurs react with safe profitable price improvements or liquidity withdrawals. 
Price improvements are liquidity provisions improving current quotes, and liquidity withdrawals are cancellations of limit orders. See \cite{budish2019theory}
for formal justifications for OBE.
OBE is weaker than MPE in the sense that it allows the existence of unilateral deviations that increase utility as long as they can be made unprofitable by others' reactions. This is reasonable as FBA leaves time for others to respond to one's actions\thomasdir{, and thus any reasonable player would not.}
\thomasdir{Therefore, the players in CLOB and FBA would follow the corresponding equilibrium (MPE and OBE, the strongest equilibrium respectively),
and the induced welfare losses are comparable.}

\section{DEX under CLOB and FBA}\label{sec:overview}
We now model DEXes under the CLOB and FBA. 
We then detail the events and provide the equilibrium analysis focusing on welfare loss and liquidity provision.

\subsection{Order of events}
We abstract the DEX as run by validators maintaining a secure blockchain. 
Arbitrageurs, investors, and traders submit transactions by sending messages to one or more validators, who then broadcast received messages to each other. In the following stochastic trading game $\mathcal{G}$, we capture the arrival of the three types of players and how the validators order and execute the transactions. 

\noindent
\colorbox{gray!10}{
 \begin{minipage}{0.95\linewidth}
 \medskip
 \noindent Trading game $\mathcal{G}$: Repeat the following for asset $X$ with initial value $V_0$. 
 \begin{compactenum}
 \item[\circno{a}] Arbitrageurs place orders on the exchange. 
 \item[\circno{b}] One of the following events is then triggered:
\begin{compactenum}
 \item[(1)] An investor arrives with probability $\player{i}$ and submits an order to the exchange. This may change order book quotes $(\vec{b_t}, \vec{a_t})$ at time $t$ 
 but not to the fundamental value. 
 \item[(2)] A private information event occurs with probability $\player{pr}$, resulting in a jump in fundamental value, and an informed trader submits an order to the exchange. This may change the order book quotes. 
 \item[(3)] A public information event occurs with probability $\player{pb}$, resulting in a jump in fundamental value, and arbitrageurs can submit withdrawals of existing orders, place new orders, or front-run stale quotes. 
 \item[(4)] Null event. The fundamental value and order books do not change.
 \end{compactenum}
 \item[\circno{c}] Each validator accumulates transactions and steps \circno{a} and \circno{b} are repeated until the validators can order transactions and output a block according to the underlying blockchain protocol. 
 \item[\circno{d}] 
 \underline{\textit{CLOB}}. After outputting a block, the validators running the exchange execute the transactions sequentially in the proposed order. 

 \underline{\textit{FBA}}. Depending on the auction frequency, after outputting one or multiple blocks, the validators execute the transactions by clearing the market with a uniform price double auction: (1) the validators first gather the orders in the current batch and all previous outstanding orders; (2) they aggregate the bids and asks; (3) if there are intersections, the market clears where supply meets demand, at a uniform price. When there are conflicting ties, priority fees lift the execution order of transactions, and further ties are broken uniformly at random. Unmatched orders that are not canceled enter into the next trading game as outstanding orders. 
 \end{compactenum}
\end{minipage}
}

\subsection{MPE under CLOB}\label{sec:clob}
As described before, we aim to solve for the markup of LPs. It is contingent on the profits from satisfying investors' orders and the costs of defending against adverse selection. 
We denote the expected fundamental value of asset $X$ conditioned on history and a new sell or buy order respectively as $\mathbb{E} [V_t | \mathcal{H}_t, buy]$ and $\mathbb{E} [V_t | \mathcal{H}_t, sell]$. 
For clarity, without loss of generality, we let each level of the book be one unit order. 
Let $Q$ be the maximum number of units investors need and $p_j$ be the probability of an investor transacting $j$ units ($j=1,\ldots, Q$). We state the results for the quotes on the limit order book (LOB) under CLOB as follows.

    \begin{theorem}
        There exists a stationary MPE in trading game $\mathcal{G}$ under CLOB processing. Bid and ask prices at the $k$-th level of the LOB in equilibrium satisfy 
        \[ b_t^k=\mathbb{E} [V_t | \mathcal{H}_t, sell] = P_t - \frac{s_{k}}{2}, a_t^k=\mathbb{E} [V_t | \mathcal{H}_t, buy] = P_t + \frac{s_{k}}{2} \] 
        where $s_{k}$ satisfies 
        \small
        \begin{multline*}
        \player{i} \sum_{j=k}^Q p_j \frac{s_{k}}{2} - \player{pr} (1 + p^{\star} \frac{\vec{\mathsf{g}}_{k} }{\vec{\mathsf{g}}_{k} (r-2) + 1 } ) \bar{J}_k \\
        - \player{pb} \bar{J}_k +(\player{i} + \player{pr}) \vec{\mathsf{g}}_{k} \mathsf{F} = 0 
        \end{multline*}
        \normalsize
       Here, $\bar{J}_k = \mathbb{P} [J>\frac{s_{k}}{2}] \mathbb{E} [J-\frac{s_{k}}{2}|J>\frac{s_{k}}{2}]$, and $\vec{\mathsf{g}}_{k}$ is the probability that an arbitrageur front-runs traders and investors at the $k$-th level. In equilibrium, $\vec{\mathsf{g}}_{k}$ is then updated to take the value that maximizes profits from front-running traders and investors, i.e., $\player{pr} p^{\star} \frac{\vec{\mathsf{g}}_{k} }{\vec{\mathsf{g}}_{k} (r-2) + 1 } \bar{J}_k - (\player{i} + \player{pr}) \vec{\mathsf{g}}_{k} \mathsf{F} $. 
       The expected price impact of an incoming $k$-unit order is $\Delta_k = \tilde{J}_k \frac{\player{pr}}{\player{pr} + \player{i}}$ with $\tilde{J}_k = \mathbb{P} [J>\frac{s_{k}}{2}] \mathbb{E} [J|J>\frac{s_{k}}{2}] $; the expected markup from liquidity providers is $(\player{pr} + \player{i})(\frac{s_{k}}{2}-\Delta_k)$. \label{thm:clob}
       \end{theorem}

The key to the proof is that in equilibrium, market-making and front-running strategies yield the same profits because arbitrageurs can choose strategies freely. 
\begin{proof}
    Let $s_{k}$ denote the spread at the $k$-th level. 
    In a DEX, unlike in centralized exchanges, an arbitrageur can also front-run privately informed traders. Since a transaction may be submitted by a common investor or privately informed trader, the front-runner can expect premiums when privately informed traders arrive. This happens with probability $\frac{\player{pr}}{\player{pr} + \player{i}}$ if the arbitrageur always front-runs ($\vec{\mathsf{g}} = \vec{1}$). More meaningfully, let the arbitrageur front-run at the $k$-th level with probability $\vec{\mathsf{g}}_{k}\in [0,1]$. As described, let the front-running transaction beat a user's transaction with probability $p^*$. 

    Suppose in an incoming order, there are $q\leq Q$ ($Q\in N$) quantities needed. We need the $k$-th spread $s_k$ ($k=1,\ldots, Q$) to satisfy that the market-making strategy yields the same returns as the front-running strategy in equilibrium:
    \small
    \begin{multline}\label{eq:spread}
        \overbrace{\player{i} \sum_{j=k}^{Q} p_j \frac{s_k}{2}}^{\text{Profits from investor}} -
        \overbrace{ \player{pr} \bar{J}_k }^{\text{Private info-induced loss}} - 
        \overbrace{(\player{pb} \frac{r-1}{r} \bar{J}_k +
        \player{pb} \mathsf{F} )}^{\text{Public info-induced loss}} = \\
        \underbrace{\player{pb} \frac{1}{r} \bar{J}_k - 
        \player{pb} \mathsf{F} }_{\text{Profits from LPs}}  + 
        \underbrace{\player{pr} p^* \frac{\vec{\mathsf{g}}_{k} }{\vec{\mathsf{g}}_{k} (r-2) + 1 } \bar{J}_k - 
        (\player{i} + \player{pr}) \vec{\mathsf{g}}_{k} \mathsf{F} }_{\text{Profits from front-running investors and traders}} 
    \end{multline}
    \normalsize
    where $\bar{J}_k = \mathbb{P} [J>\frac{s_{k}}{2}] \mathbb{E} [J-\frac{s_{k}}{2}|J>\frac{s_{k}}{2}] $ is the expected extra value change uncovered by the spread (induced by the asset fundamental value jump), and $p_j$ is the probability of an investor transacting $j$ units. 
    
    In \Cref{eq:spread}, the left-hand side is the profits from the market-making strategy and the right-hand side is the returns from front-running LPs and private traders. 
    To update $\vec{\mathsf{g}}_{k}$ stored in the current state, we lay out the right-hand side as a function of $\vec{\mathsf{g}}_{k}$ and solve for $\vec{\mathsf{g}}_{k}$ that maximizes the function value in range $[0,1]$ for a front-running arbitrageur, by taking the first and second derivatives of the function. 
    One can solve it analytically or numerically if closed-form formulas do not exist for specific distributions of the jump $J$. 

    We now compute the additional transaction costs for common investors. When an arbitrageur front-runs common investors, the fundamental value does not change. This front-runner only earns premiums when private traders arrive, which happens with probability $\frac{\player{pr}}{\player{pr} + \player{i}} $. 
    To compute the welfare loss for investors, we can then treat front-running orders as part of private traders' orders. Depending on the size of the jump, they can place multi-unit orders profitably, and we first consider a single unit. Let $\Delta_1$ be the expected price impact of an incoming unit order at time $t$. 
    Since one cannot distinguish between orders from informed traders from investors, we have 
    \small
    \begin{align*}
        \Delta_1 &= \pm \mathbb{E} [V_t | \mathcal{H}_t, \player{i} \vee \player{pr}] \mp P_t \\
        &= \pm \mathbb{E} [V_t | \mathcal{H}_t, \player{i} \vee \player{pr}] \mp \mathbb{E} [V_t | \mathcal{H}_t] = \tilde{J}_1 \frac{\player{pr}}{\player{pr} + \player{i}} 
    \end{align*}
    \normalsize
    where $+,-$ indicate buy, sell order and $\tilde{J}_k = \mathbb{P} [J>\frac{s_{k}}{2}] \mathbb{E} [J|J>\frac{s_{k}}{2}] $ (for $1\leq k\leq Q$). After absorbing the price impact, the additional markup from the liquidity provider on the first level of the order book is then $M_{C} = \frac{s_{1}}{2}-\Delta_1$. 
    The expected markup is then $(\player{pr} + \player{i})M_{C}$. We can calculate multi-unit orders in the same way. Let $\Delta_k$ be the expected price impact of the $k$-unit order. We can compute $\Delta_k= \tilde{J}_k \frac{\player{pr}}{\player{pr} + \player{i}} $. Then the markup on the $k$-th level of the order book is then $M_{C}^{(k)} = \frac{s_{k}}{2}-\Delta_k$, and its expected value is $(\player{pr} + \player{i}) M_{C}^{(k)}$. 
\end{proof}

\vspace{1mm}
\noindent
\colorbox{gray!15}{
 \begin{minipage}{0.95\linewidth}
    \paragraph{Interpret \Cref{thm:clob}}
    The spread $s$ at each level of the order book consists of (1) profits for market-making and (2) the part to absorb the price impact $\Delta$ from orders driven by private information. 
    The price impact exists because arbitrageurs cannot distinguish common trades that do not carry information (which affects asset fundamental values) from privately informed trades. 
    The expected welfare loss of investors and traders is contained in the

    profits, i.e., $(\player{pr} + \player{i})(\frac{s}{2}-\Delta)$, which we address as markups. 
    They are affected by the arrival rates of investors ($\player{i}$), private information ($\player{pr}$), and \thomasdir{public information} ($\player{pb}$), the front-running probability ($\vec{\mathsf{g}}$) and its success probability ($p^{\star}$), priority fees $\mathsf{F} $, jump size, the number of arbitrageurs ($r$), and transaction sizes. 
    \thomasdir{Below, we discuss how the markup within the spread is affected intuitively. 
    In \Cref{sec: empirical}, we show how these formulas are reflected using real-life data, and in \Cref{sec: example}, we give an example demonstrating this theorem.}
\end{minipage}
}

\medskip
\paragraph{What affects the markups under CLOB}
From the theorem, overall, the markup from LPs in equilibrium increases with the jump size, public information releases, the front-running probability, and its success probability. It decreases with private information arrivals, fees, and the number of arbitrageurs.
Fix a specific front-running probability $\vec{\mathsf{g}}$, the markup first increases as the investor arrival rate rises and then decreases as it continues to grow. This is because $\player{i}$ raises market-making profits from common investors. In the meantime, it also lowers the profits from front-running traders due to increased costs (i.e., front-running arbitrageurs attack unprofitable common trades more often). Since arbitrageurs can choose freely between market-making and front-running, more arbitrageurs adopt the market-making strategy when market-making is more profitable. The competition reduces the markup until equilibrium is reached, where the profits from both strategies are equal.

\subsection{OBE under FBA}\label{sec: obe}
Under FBA, we only need to process excessive demand or supply since only those transactions need to be fulfilled by LPs. 
We assume the withdrawals and updates of existing quotes can be completed in time during the batch auction interval. 
Let $Z_I$ denote the excess demand (demand minus supply) during a batch interval $I$. $Z_I\leq Q+1$ ($Q\geq 0$ is an integer) is bounded. 
We borrow $\bar{J}_k$ and $\tilde{J}_k$ from \cref{thm:clob} and formally state the result for FBA in \cref{thm: fba}. 

    \begin{theorem}
        There exists an OBE in pure strategies in trading game $\mathcal{G}$ under FBA processing. If $|Z_I | \leq Q + 1$, bid and ask quotes in equilibrium satisfy 
        \[
        \begin{cases}
            b^k_{(I)} = P_I - \Delta_k - M_k & Z_I <0\\
            a^k_{(I)} = P_I + \Delta_k + M_k & Z_I > 0\\
        \end{cases}
        , \ k=1,\ldots, Q+1 
        \]
        where $\Delta_k = \tilde{J}_k \frac{\player{pr}}{\player{pr} + \player{i}}$ is the expected price impact from informed traders, $M_k = \sum_{u=k}^Q \Delta_u \prod_{v=k}^{u} \alpha_{v}$ for $k=1,\ldots, Q$,
        and $M_{Q+1} = 0$ are the markups from LPs, $\alpha_k = \frac{q_{k+1}}{q_k + q_{k+1}}$ and $q_k = \mathbb{P} [Z_I = k |\ | Z_I|\leq Q+1]$. 
        The expected markup per unit time is $\frac{2}{I} \sum_{k=1}^Q k q_k M_k$. \label{thm: fba}
       \end{theorem}

The proof utilizes induction on $k$. Intuitively, first, the ask quote for the $(Q+1)$-th level LOB has $0$ markup as otherwise, another arbitrageur can improve on it to settle the $(Q+1)$-th transaction. This means that the last quote reflects only the price impact. In the induction process, given the \mbox{$(k+1)$-th} quote ($k\leq Q$), we can determine the $k$-th markup by considering safe price improvements or other deviations in different cases of $Z_I$, which can take values from $k$ (with probability $q_k$), $k+1$ (with probability $q_{k+1}$), to $Q+1$ (with probability $q_{Q+1}$), and different cases of an LP, who may already own a lower-level quote, an upper-level quote, or no existing quote.
\begin{proof}
    We show by induction that the markups $\{M_k\}_{k=[Q+1]}$ are the largest such that there exist (i) neither strictly profitable safe price improvements (ii) nor strictly profitable robust deviations. More specifically, (i) means that for a targeted quote at the $k$-th level, an arbitrageur cannot safely insert a new profitable quote at this level, i.e., pushing the original $k$-th quote to be the new $(k+1)$-th quote and so forth, and the current $k$-th quote issuer cannot improve this quote (after being pushed) to safely increase profits. (ii) means that the $k$-th quote issuer cannot robustly increase profits by issuing or updating quotes deviating from the stated markups, considering that others can react with quote withdrawals and updates. We will focus on the ask quotes, but we only need to adapt signs for analyzing bid quotes. 
   
    \paragraph{$(Q+1)$-th level} We first show that $M_{Q+1} = 0$ is the OBE, i.e., $a^{Q+1}_{(I)} = P_I + \Delta_{Q+1}$. We write $a^{Q+1}_{(I)}$ as $a^{Q+1}$ in the following analysis for visual simplicity. (i) We show that safe profitable price improvements on the last quote with zero markups do not exist. First, an arbitrageur inserting a new $(Q+1)$-th quote with a price lower than $a^{Q+1}$ experiences a loss in expectation. Because the spread does not cover the potential jump in asset value. Similarly, the current $(Q+1)$-th quote issuer does not have the incentive to lower the quote. Second, the issuer is not incentivized to increase the quote either, i.e., to $\bar{a}^{Q+1} = P_I + \Delta_{Q+1}+ \epsilon$ for even a small $\epsilon>0$. Because this is not a safe price improvement: another arbitrageur can safely insert a quote at price $a^{Q+1}$ and push this issuer's new quote to be the $(Q+2)$-th quote. Since $Z_I\leq Q+1$, the expected profit after others' reaction is zero, which is less than the original profit. 
    (ii) With similar reasoning, we can show that robust profitable deviation does not exist. This is because the owner of the $(Q+1)$-th quote is not incentivized to insert a new quote or update another quote to be the $(Q+1)$-th that deviates from zero markups. 
   
    \paragraph{$(k+1)$-th level} Suppose there do not exist safe profitable price improvements or robust profitable deviations on the $(k+1)$-th quote. The markups on the $(k+1)$-th level and onwards to the $(Q+1)$-th level are as stated in \Cref{thm: fba}. 
   
    \paragraph{$k$-th level} We show that the stated $M_{k}$ is the OBE given the quotes at the $(k+1)$-th level. (i) Suppose there exists a safe profitable price improvement on the $k$-th quote by $\epsilon$, i.e., $\bar{a}^{k} = a^{k} - \epsilon$. First, consider an arbitrageur who does not own quotes and inserts a new $k$-th quote. If $Z_I= k$, then the arbitrageur expects positive profits, and if $Z_I > k$, then a loss is expected because the settlement price would be below $\bar{a}^{k}$. Let $\Delta_k = \tilde{J}_k \frac{\player{pr}}{\player{pr} + \player{i}}$ and $q_k = \mathbb{P} [Z_I = k |\ | Z_I|\leq Q+1]$. We can express the profits as:
    \small
    \[ A_k = q_k (M_k - \epsilon) + \sum_{i=k+1}^{Q+1} q_i (M_{i-1} - \Delta_{i-1}) 
    \]
    \normalsize
    For this improvement to be safe, the largest markup needs to be achieved, i.e., $\epsilon\rightarrow 0$, so that no other arbitrageurs have the incentive to undercut again. We next demonstrate the conditions such that $A_k < 0$. 
    We know that the analog equation for the $(k+1)$-th level ask quote is 
    \small
    \[ A_{k+1} = q_{k+1} (M_{k+1} - \epsilon) + \sum_{i=k+2}^{Q+1} q_i (M_{i-1} - \Delta_{i-1}) < 0 \] 
    \normalsize
    Observe that instead of directly requiring $A_k < 0$, we can locate conditions that let $A_k = A_{k+1}$ as $\epsilon\rightarrow 0$. This gives us $M_{k} = \alpha_k \Delta_k +\alpha_k M_{k+1}$ where $\alpha_k = \frac{q_{k+1}}{q_k + q_{k+1}}$. 
    Applying this recursively, we obtain $M_k = \sum_{u=k}^Q \Delta_u \prod_{v=k}^{u} \alpha_{v}$. In other words, when the original markup of the $k$-th level quote satisfies this condition, $A_k=A_{k+1}<0$. 
   
    Now consider an arbitrageur who owns some existing quotes. To implement price improvements, she can insert a new $k$-th quote or adjust existing quotes. As discussed above, insertion results in a loss. 
    If she moves up a lower quote at the $k^*$-th ($k^* < k$) level to the $k$-th, the expected profit is lower because some profits are lost for $k^* \leq Z_I < k$. If she moves down one of the quotes to be the new $k$-th quote, then the extra profit generated is bounded by $A_k$. Because $A_k$ is the extra profits of an arbitrageur who originally expects $0$ returns. 
    
    (ii) Suppose there exists a robust profitable deviation for the issuer of the $k$-th quote. To allow the issuer to deviate other than improving quotes (which we just discussed), we let the issuer own some quotes under the $k$-th level. The issuer can update the $k$-th quote to be the $k'$-th quote ($k'\geq k$) with a markup higher than the original $M_{k'}$ but still lower than $\Delta_{k'+1} + M_{k'+1}$. This is profitable but not robust: another arbitrageur can re-issue the original $k$-th level quote and push the updated $k'$-th quote to be the $(k'+1)$-th. This arbitrageur's price improvement is safe as it restores the original $k$-th quote. 
   
    By induction, the markups from LPs at each level satisfy the established conditions. 
    The expected markup of the liquidity providers per unit of time can then be calculated as $\frac{2}{I} \sum_{k=1}^Q k q_k M_k$. 
\end{proof}

\vspace{1mm}
\noindent
\colorbox{gray!15}{
 \begin{minipage}{0.965\linewidth}
 \paragraph{Interpret \Cref{thm: fba}}
 Same as before, the spread consists of the markup and the cushion for absorbing the price impact from adverse selection. The expected markup or welfare loss is influenced by the arrival rates of investors ($\player{i}$), privately informed traders ($\player{pr}$), the jump size, and the excessive demand. 
  \thomasdir{Again, we discuss these relationships below, show real-world data-based analysis in \Cref{sec: empirical}, and give an example in \Cref{sec: example}.}
 \end{minipage}
}
\vspace{1mm}

\paragraph{What affects the markups under FBA}
Overall, the welfare loss in equilibrium increases with the jump size, private information releases, and excessive demand, and decreases with investor arrivals. First, when the jump size or private information arrival rates are larger or the investor arrival rates are smaller, the adverse selection is more severe, and the expected price impact rises. This in turn results in higher markups. Second, when there is more excessive demand (or supply), then the markups on the ask (or bid) quotes grow to incorporate more levels of price impacts on the LOB.

\section{Welfare loss and liquidity provision}\label{sec:comparison}
We compare the welfare loss and liquidity provision under CLOB and FBA in equilibrium in \cref{sec: welfare comparison} and \cref{sec: liquidity comparison},
\thomasdir{give an empirical analysis with real world data in \cref{sec: empirical},}
and show an example with small parameters in \cref{sec: example}.

\subsection{Welfare loss}\label{sec: welfare comparison}
This section compares FBA and CLOB under special cases and general conditions and ends with the main takeaways.

\begin{figure*}[!ht]
    \centering
    \begin{subfigure}[b]{0.48\textwidth}
    \centering
    \includegraphics[width=\textwidth]{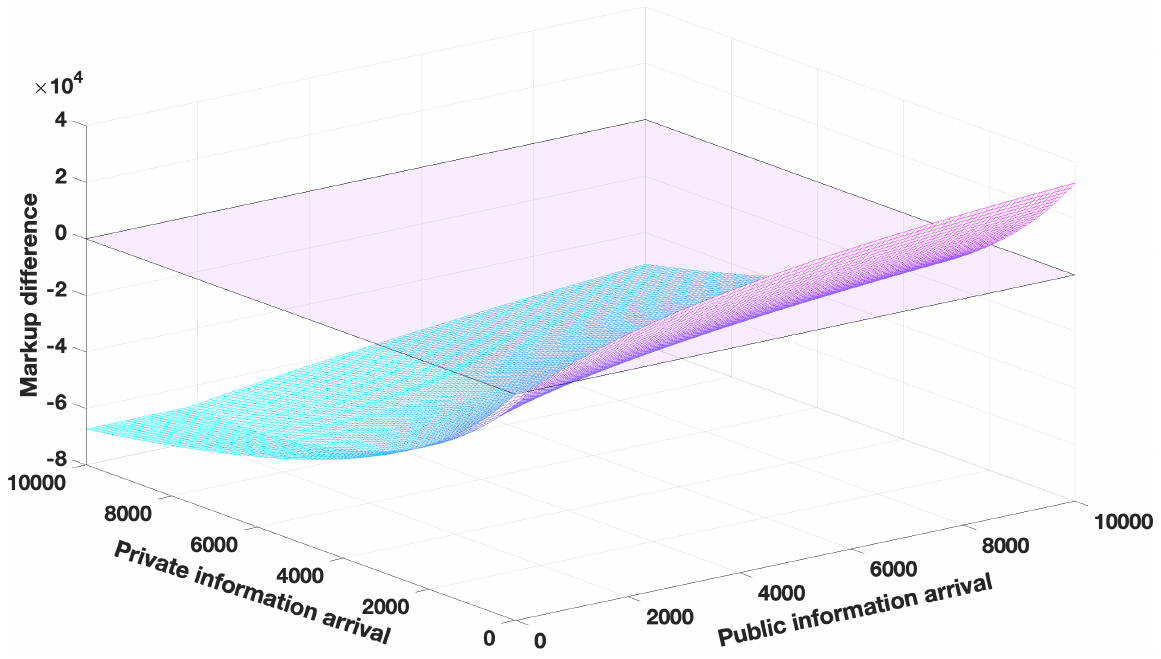}
    \caption{Bad case for both CLOB and FBA.}
    \label{fig:pbpr bb}
    \end{subfigure}
    \hfill
    \begin{subfigure}[b]{0.48\textwidth}
    \centering
    \includegraphics[width=\textwidth]{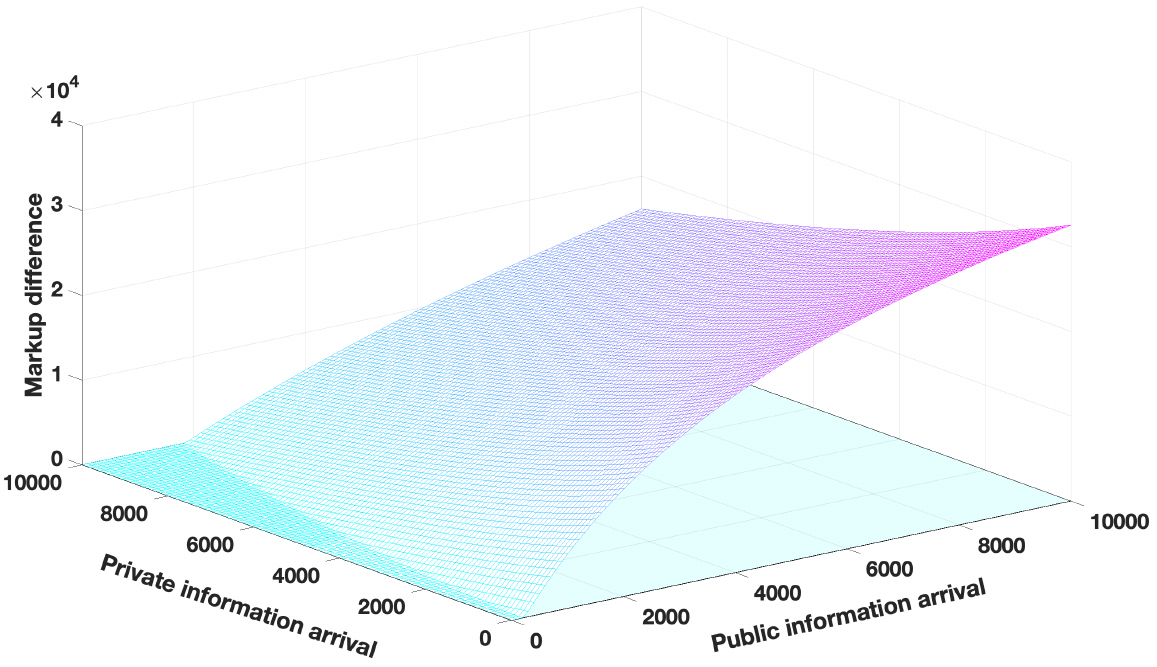}
    \caption{Bad case for CLOB and good case for FBA.}
    \label{fig:pbpr bg}
    \end{subfigure}
    \hfill
    \begin{subfigure}[b]{0.48\textwidth}
    \centering
    \includegraphics[width=\textwidth]{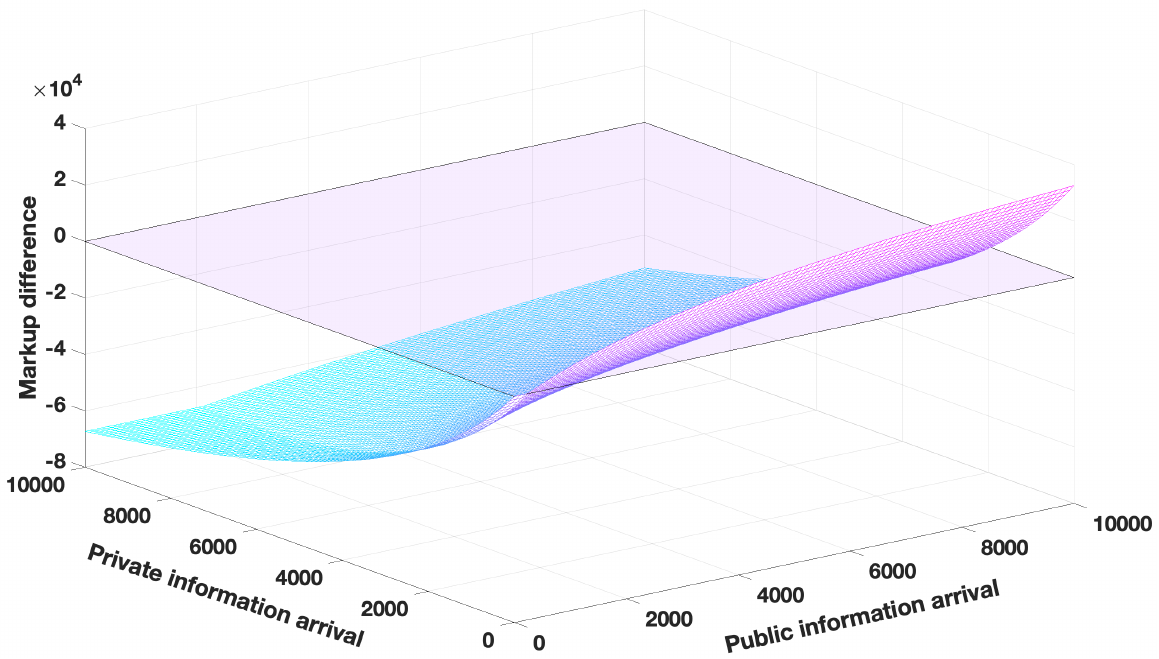}
    \caption{Good case for CLOB and bad case for FBA.}
    \label{fig:pbpr gb}
    \end{subfigure}
    \hfill
    \begin{subfigure}[b]{0.48\textwidth}
    \centering
    \includegraphics[width=\textwidth]{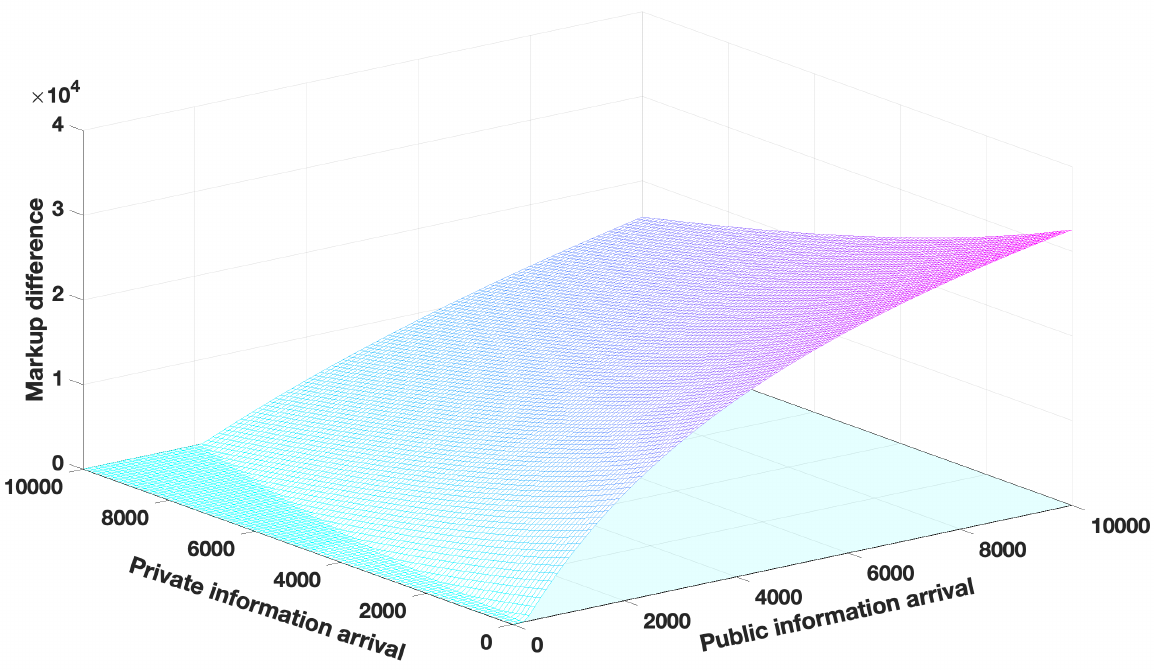}
    \caption{Good case for both CLOB and FBA.}
    \label{fig:pbpr gg}
    \end{subfigure}
    \caption{The markup difference between CLOB and FBA with respect to public and private information arrivals. Positive regions are where FBA has fewer markups, i.e., less welfare loss. The investor arrival rate $\player{i}$ is set to $5000$. Increasing (decreasing) $\player{i}$ pushes the surface up (down).
    }
    \label{fig:pbpr}
    \begin{subfigure}[b]{0.48\textwidth}
    \centering
    \includegraphics[width=\textwidth]{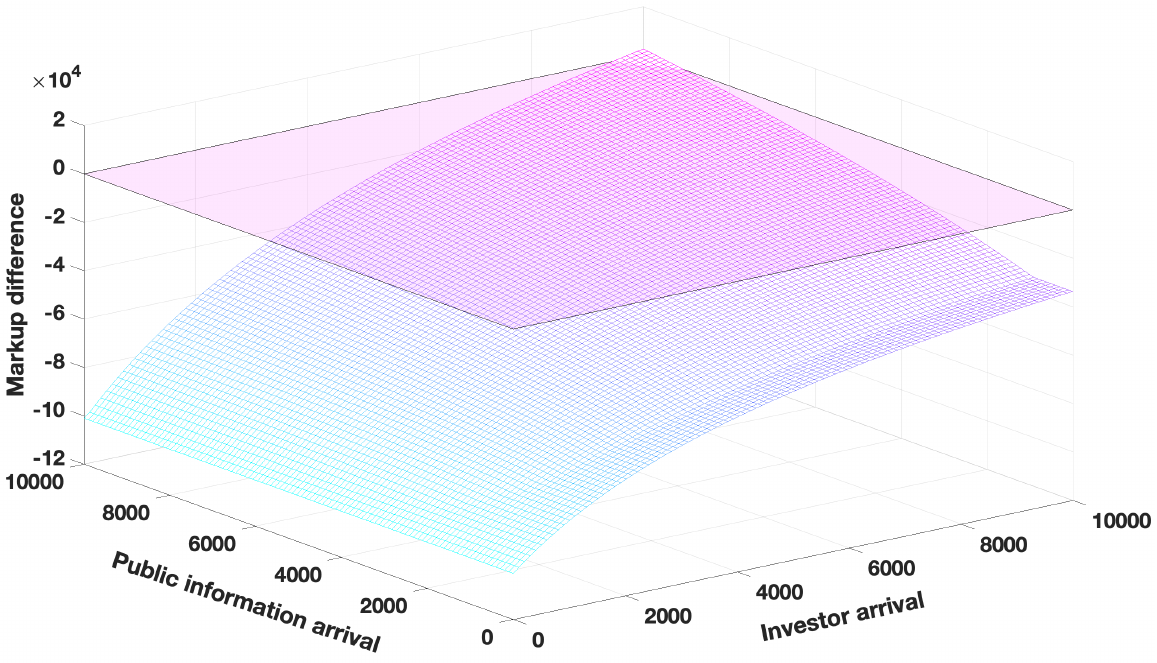}
    \caption{Bad case for both CLOB and FBA.}
    \label{fig:ipb bb}
    \end{subfigure}
    \hfill
    \begin{subfigure}[b]{0.48\textwidth}
    \centering
    \includegraphics[width=\textwidth]{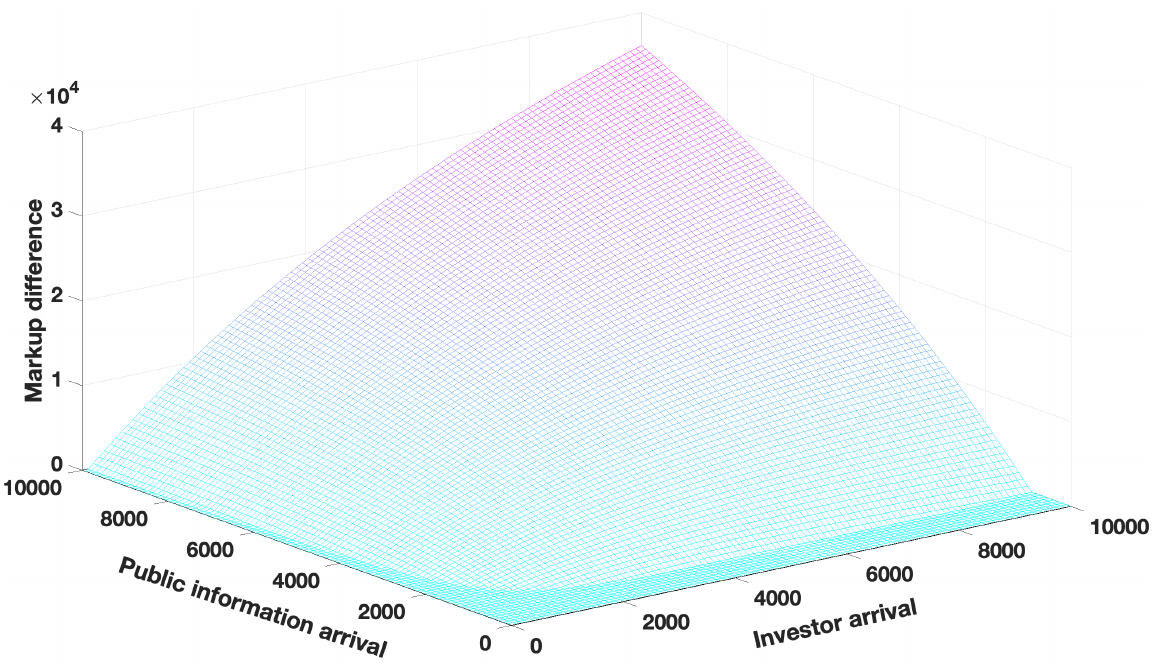}
    \caption{Bad case for CLOB and good case for FBA.}
    \label{fig:ipb bg}
    \end{subfigure}
    \hfill
    \begin{subfigure}[b]{0.48\textwidth}
    \centering
    \includegraphics[width=\textwidth]{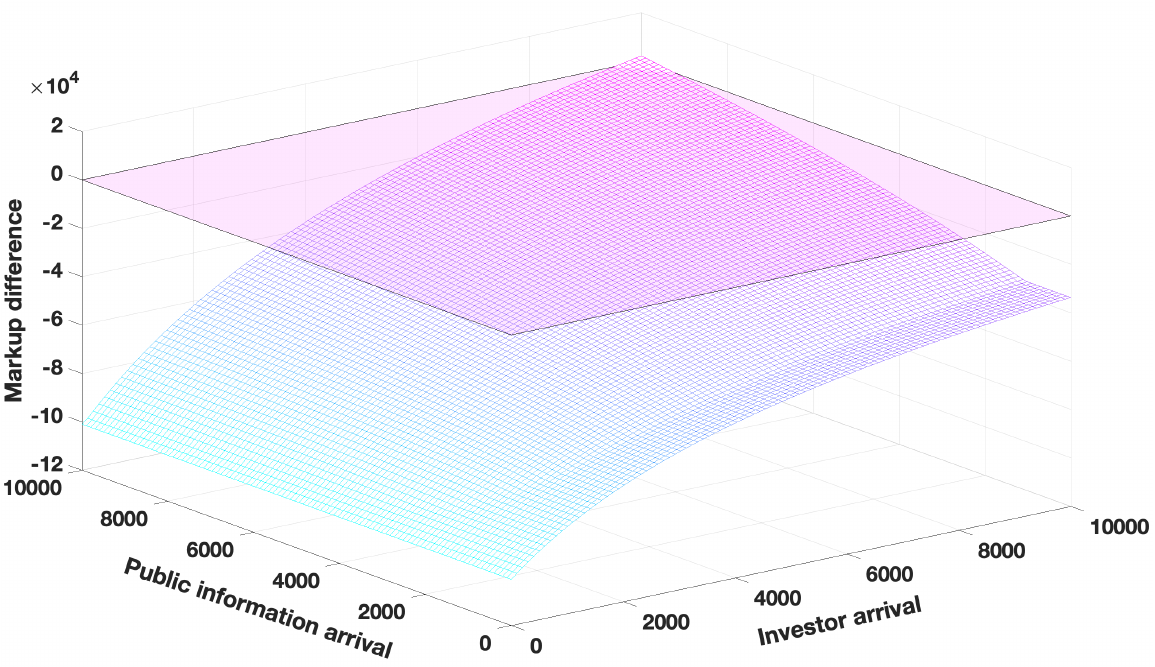}
    \caption{Good case for CLOB and bad case for FBA.}
    \label{fig:ipb gb}
    \end{subfigure}
    \hfill
    \begin{subfigure}[b]{0.48\textwidth}
    \centering
    \includegraphics[width=\textwidth]{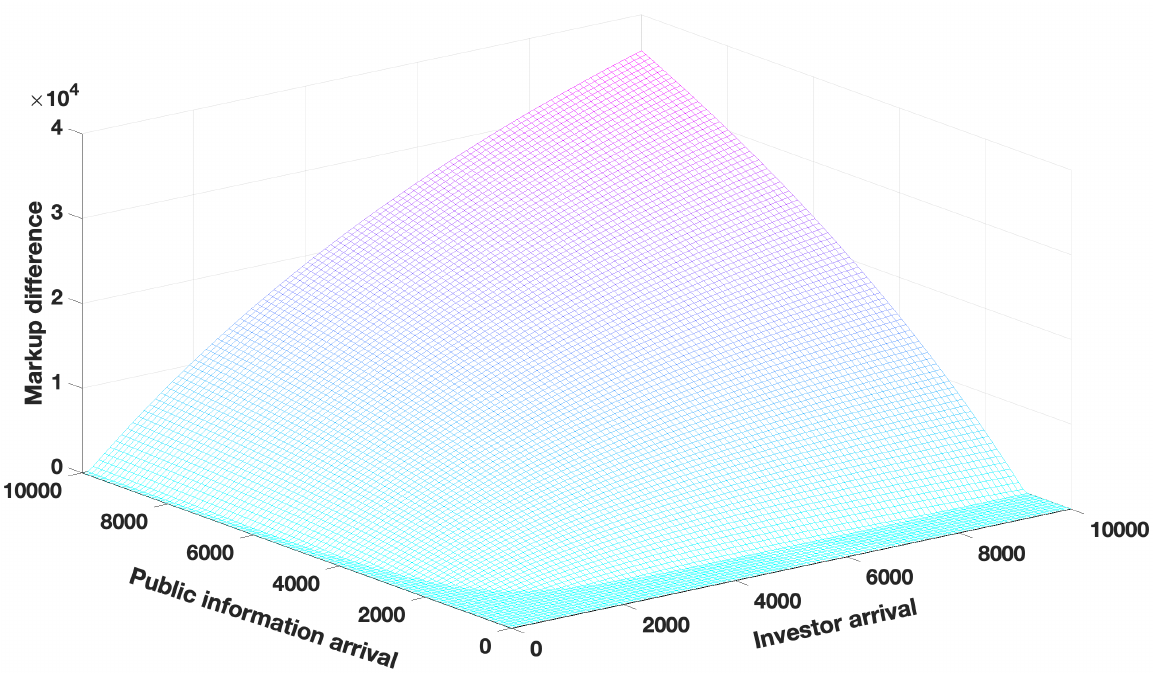}
    \caption{Good case for both CLOB and FBA.}
    \label{fig:ipb gg}
    \end{subfigure}
    \caption{The markup difference between CLOB and FBA with respect to public information and investor arrivals. The private information arrival rate $\player{pr}$ is set to be $5000$. Increasing (decreasing) $\player{pb}$ pushes the surface down (up).
    }
    \label{fig:ipb}
   \end{figure*}

\begin{figure*}[!th]
       \centering
       \begin{subfigure}[b]{0.48\textwidth}
       \centering
       \includegraphics[width=\textwidth]{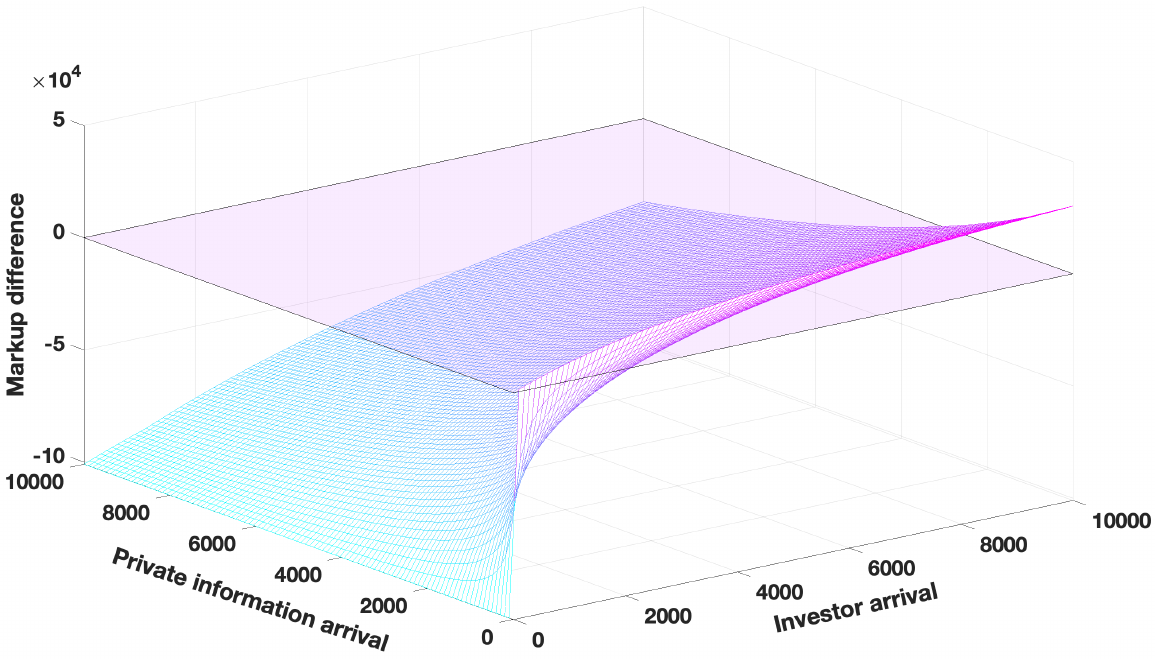}
       \caption{Bad case for both CLOB and FBA.}
       \label{fig:ipr bb}
       \end{subfigure}
       \hfill
       \begin{subfigure}[b]{0.48\textwidth}
       \centering
       \includegraphics[width=\textwidth]{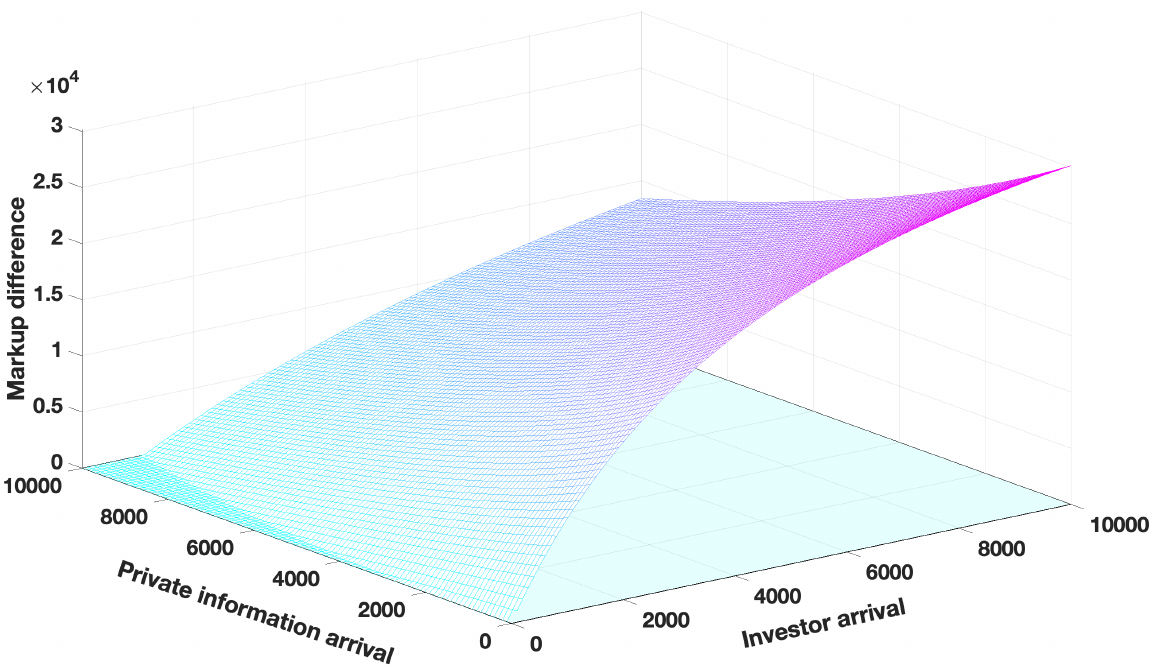}
       \caption{Bad case for CLOB and good case for FBA.}
       \label{fig:ipr bg}
       \end{subfigure}
       \begin{subfigure}[b]{0.48\textwidth}
       \centering
       \includegraphics[width=\textwidth]{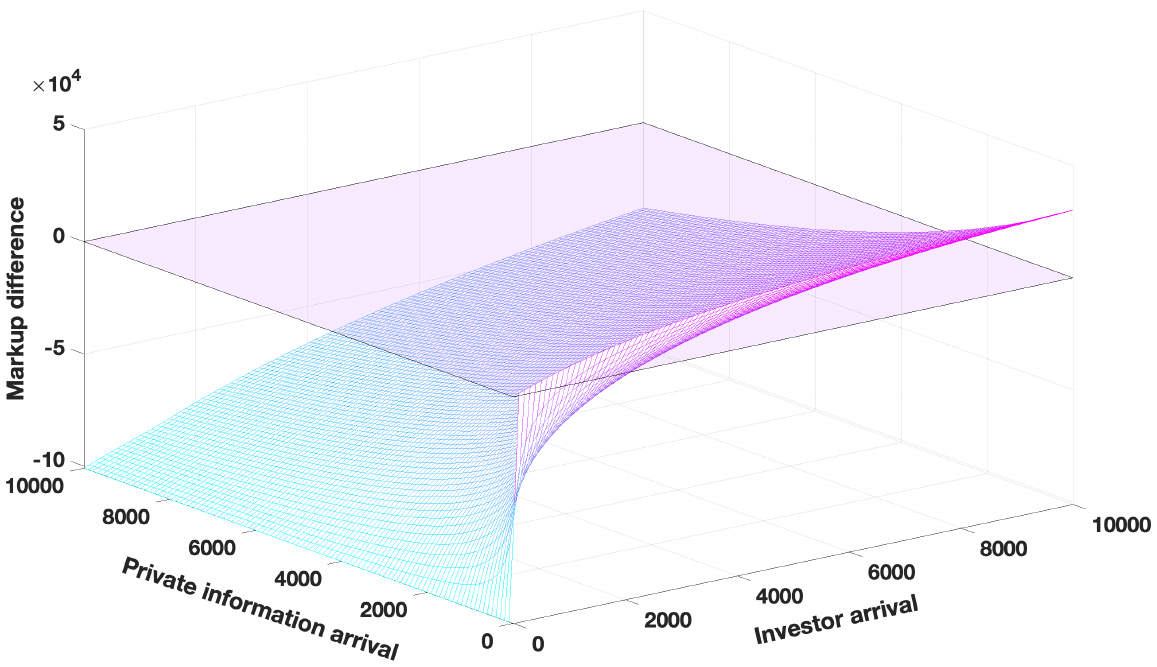}
       \caption{Good case for CLOB and bad case for FBA.}
       \label{fig:ipr gb}
       \end{subfigure}
       \hfill
       \begin{subfigure}[b]{0.48\textwidth}
       \centering
       \includegraphics[width=\textwidth]{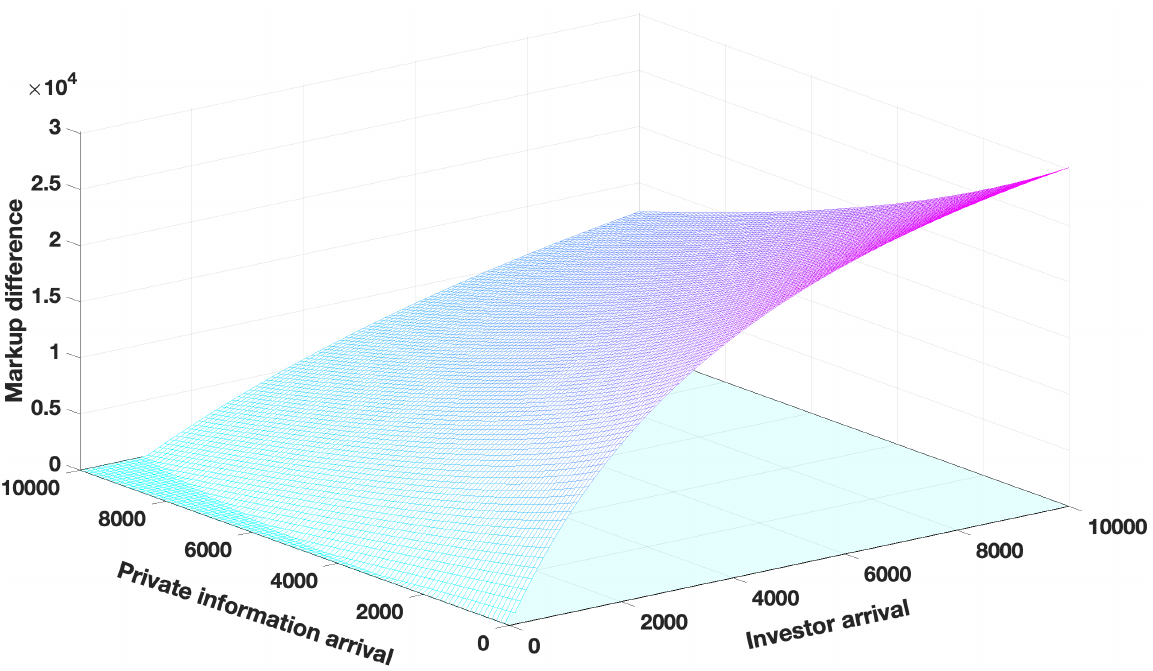}
       \caption{Good case for both CLOB and FBA.}
       \label{fig:ipr gg}
       \end{subfigure}
       \caption{The markup difference between CLOB and FBA with respect to private information and investor arrivals. The public information arrival rate $\player{pb}$ is set to be $5000$. Increasing (decreasing) $\player{pb}$ pushes the surface up (down).
       }
       \label{fig:ipr}
   \end{figure*}

\paragraph{Special cases}
When there are no privately informed traders ($\player{pr} = 0$) but there exists public information ($\player{pb} > 0$), FBA is strictly better than CLOB because the markup in FBA becomes zero while the markup in CLOB is still positive. As depicted in \cref{fig:pbpr} and \ref{fig:ipr}, FBA always suffers from less welfare loss when $\player{pr}=0$. When there is no public information 
   \FloatBarrier
\noindent
($\player{pb} = 0$) but there exists private information ($\player{pr} > 0$), FBA has positive markups that increase with $\player{pr}$. For CLOB, the markups tend to zero for small $\player{pr}$ but remain positive and also increase with $\player{pr}$ otherwise. This is also portrayed in \cref{fig:pbpr} where the markups for both processing models approach zero.

\paragraph{General comparison}
The parameters affecting markups in both models are $\player{i}, \player{pr}, \player{pb}$. We examine how they affect the markup differences in CLOB and FBA in four different settings (a)-(d). 
In setting (a), we pick the bad case for the remaining effective parameters for both CLOB and FBA. 
This means a high front-running success probability, small priority fees for CLOB, and high excessive demand for FBA. Parameters in settings (b)-(d) are set similarly to reflect the combinations of good- and bad-case scenarios for the two models. 
Additionally, in FBA, the excess demand $Z_I$ follows a truncated Skellam distribution as the arrival rates of investors and traders follow the compound Poisson jump process. Therefore, when computing the markup differences, we consider the upper bound for the markup in FBA. This means that FBA can perform \textit{strictly better} than the predictions. 

\underline{Public versus private information.} 
Overall, as shown in \cref{fig:pbpr}, more public information renders FBA appealing \thomasdir{since markups in CLOB increase} due to front-running. As a result, FBA performs better when $\player{pb}$ is high compared to $\player{pr}$. 
More private information gives CLOB an advantage by increasing the spread in CLOB. 
As for the state of the market, if it is one-sided and there is a large excessive demand, CLOB outperforms FBA and FBA realizes lower welfare loss only when $\player{pr}$ is small. 
When the market is balanced and submitted orders are mostly settled among themselves, FBA has lower welfare loss when $\player{pb}$ surpasses $\player{pr}$ by a smaller amount. 
This can also be observed in \cref{fig:ipb} and \ref{fig:ipr}: in the good cases for FBA where the market is more balanced, FBA outperforms CLOB in the majority of the parameter regions. 

\underline{Investor arrivals versus public information.} 
We know from \cref{thm: fba} that the markups for FBA do not depend on public information while markups under CLOB increase with $\player{pb}$. 
As a result, FBA performs better in terms of welfare as $\player{pb}$ increases, which is also evident in \cref{fig:ipb}. 
Investor arrivals have a mixed effect. FBA experiences a smaller price impact as $\player{i}$ grows larger, diminishing its markup. CLOB also faces a smaller price impact. And larger $\player{i}$ decreases the profits from front-running and more arbitrageurs compete as LPs, reducing the markups. Overall, FBA realizes better social welfare if sufficiently many market participants are price takers or are publicly informed.

\underline{Investor arrivals versus private information.} 
As described in \Cref{sec: obe}, FBA's welfare loss decreases with the investor arrival rate and increases with private information. CLOB's induced welfare loss decreases with $\player{pr}$ and also eventually decreases with $\player{i}$. 
As shown in \Cref{fig:ipr}, in the bad case for FBA, its markups are amplified by the excessive demand and only larger $\player{i}$ absorbs this effect. In the good case for FBA, its markups are already small even when $\player{i}$ is minute. 
\FULL{Overall, FBA achieves better welfare when $\player{i}$ is sufficiently high and $\player{pr}$ is sufficiently small. }

\thomasdir{
\paragraph{Main takeaways}
In a nutshell,
we believe FBA is more suitable for the setting of blockchain systems,
since we believe blockchain markets tend to have (a lot) more public information than private information.
For example, for Bitcoin and Ethereum,
all the core codes and projects are open-sourced, and the transactions are overt on-chain.
Furthermore, the main goal of blockchain systems is to be decentralized and to have less hidden information.
Under this belief, FBA tend to have less welfare loss,
as also shown in \Cref{sec: empirical}.
}

\subsection{Liquidity comparisons from bid-ask spread}\label{sec: liquidity comparison}
We now turn to compare the two execution models from a liquidity provision perspective, with the bid-ask spread being the proxy quantity that we examine. The spread consists of the price impact from orders driven by information about assets' valuation changes, and markups from LPs.

We include the figures depicting how $\player{i}, \player{pb}, \player{pr}$ affect the spreads in FBA and CLOB in \Cref{fig: spread pbpr,fig: spread ipb,fig: spread ipr} (in \cref{app: liquidity comparison}). 
Similar to the analysis with welfare loss, $\player{pr}$ approaching $0$ always benefits liquidity provision in FBA. For a larger $\player{pr}$, FBA has better liquidity provision only when $\player{i}, \player{pb}$ are sufficiently high with respect to $\player{pr}$. When the market is thin or one-sided and more orders are fulfilled with LPs' orders, this requirement on $\player{i}$ and $\player{pb}$ is even more demanding. When the DEX has a more balanced market and a larger proportion of the orders can be filled among themselves, the requirement is looser. 

\thomasdir{
Therefore, similar to the previous section,
we conclude that FBA provides better liquidity in the blockchain settings,
under the assumption that the blockchain market has much more public information than private information.}

\subsection{Empirical analysis} \label{sec: empirical}
We sample 707,267 BTC-USD transactions and 786,727 ETH-USD transactions on dYdX~\cite{dydx} from January 2023 to October 2024 via Tardis~\cite{tardis}. 
We then simulate CLOB and FBA processing on the order book at the receiving time (measured in nanoseconds) of each transaction. The simulation code and result summary are available here~\cite{simulationsrc}. We compute the realized spread~\cite{indriawan2020effects} as the welfare loss. We trim outliers in the computed realized spread value with three median absolute deviations for a more robust comparison. 

\begin{figure}[!htbp]
    \centering
    \begin{subfigure}[b]{0.3\textwidth}
        \centering
        \includegraphics[width=\linewidth]{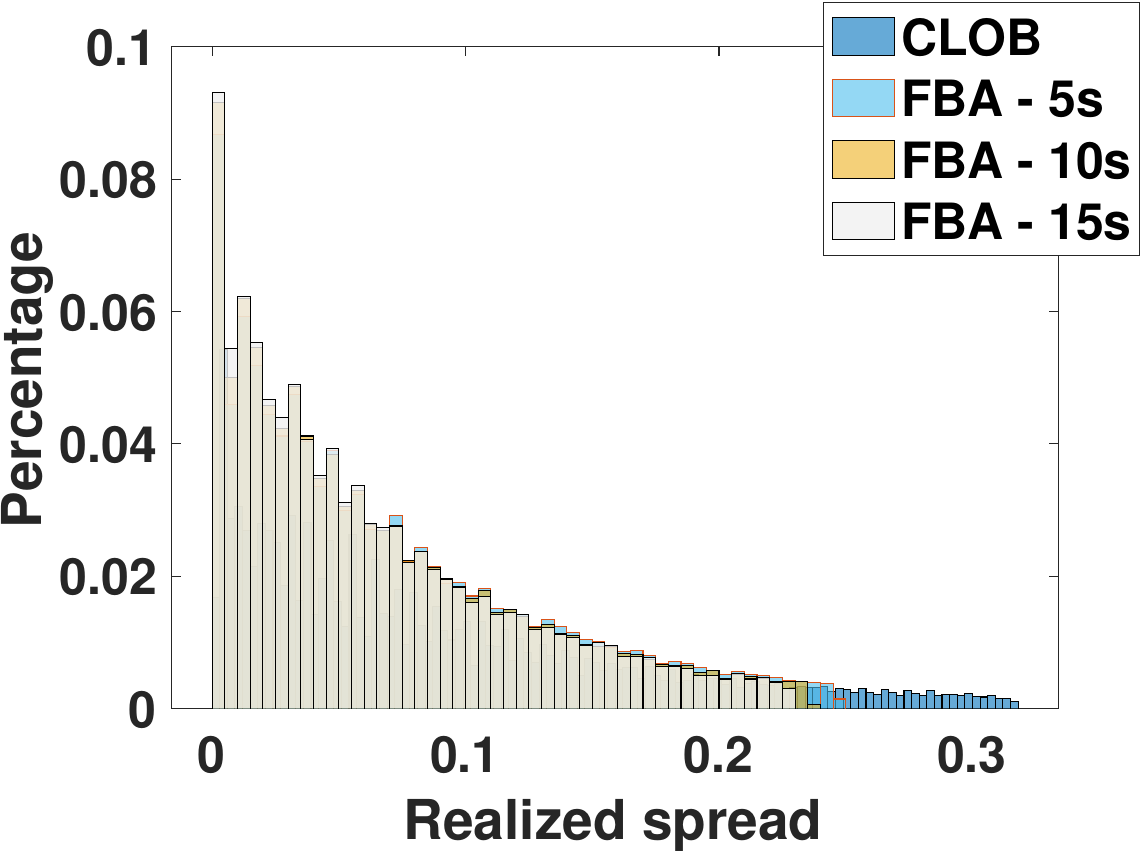}
        \caption{BTC-USD.}
        \label{fig: btc rspread}
    \end{subfigure}
    \hfill
    \begin{subfigure}[b]{0.3\textwidth}
        \centering
        \includegraphics[width=\linewidth]{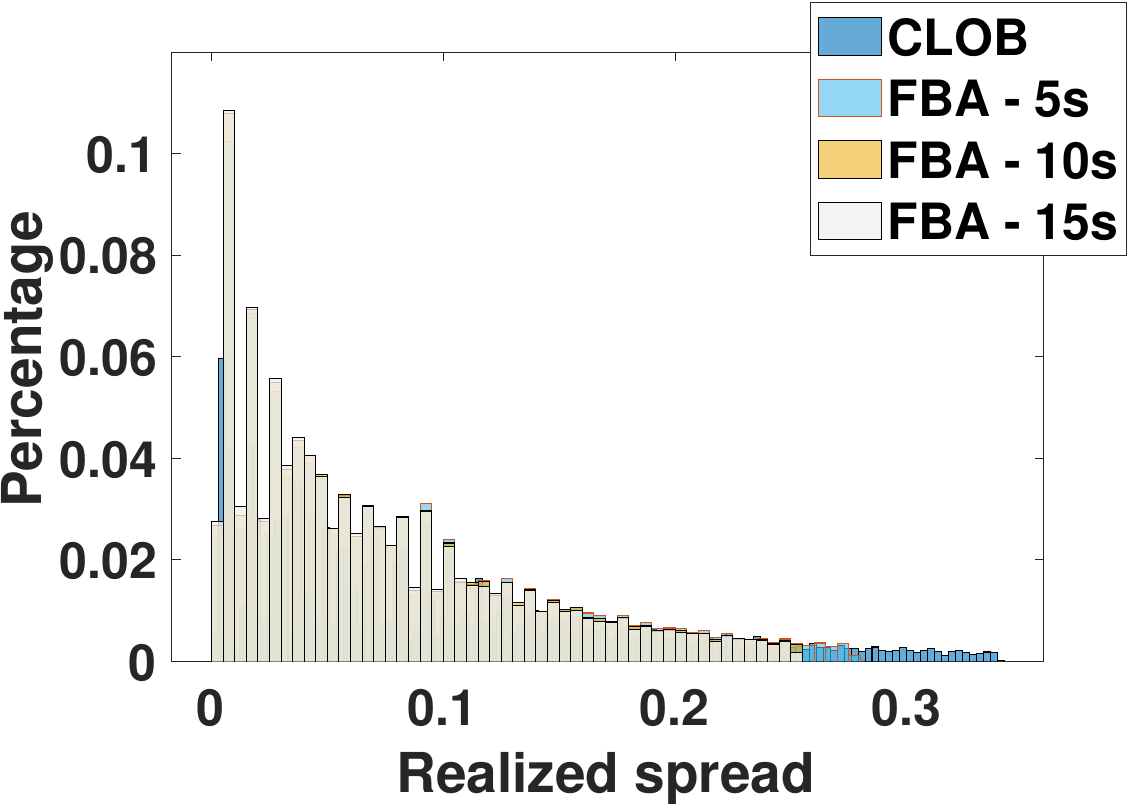}
        \caption{ETH-USD.}
        \label{fig: eth rspread}
    \end{subfigure}
    \caption{Distribution of realized spread (with outliers trimmed) from simulations on sampled transactions. \tian{The bars for depicting the four execution methods are placed on top of each other. The bars for FBA with 15 second auction frequency concentrate on the lower end, while the rest stretch to higher realized spread, with CLOB having the rightmost bars.}}
\end{figure}

For BTC-USD, the realized spread of CLOB has a mean of $0.0864$ and a median of $0.0643$; the realized spread of FBA has mean values $(0.0688, 0.0652, 0.0632)$ and median values $(0.0520, 0.0490, 0.0476)$ for auction frequency of 5, 10, and 15 seconds. CLOB inflicts $24\%$-$37\%$ more transaction costs. As shown in \Cref{fig: btc rspread}, the realized spread for FBA, especially with 5-second auction frequency is more concentrated at the smaller end. The actual clearing price is on average $1.06$, $1.74$, and $2.20$ better than the posted prices of settled trades under FBA with auction frequency of 5, 10, and 15 seconds. A longer auction period allows for the arrival of more transactions, bringing about more and at least not worse trading opportunities. 

For ETH-USD, the realized spread of CLOB has a mean value $0.0950$ and a median value $0.0710$. FBA has mean realized spreads of $(0.0770, 0.0720, 0.0709)$ and median realized spreads of $(0.0586, 0.0550, 0.0544)$ for auction frequency of 5, 10, 15 seconds. CLOB has $21\%$-$34\%$ more transaction costs. The clearing price is on average $0.05$, $0.11$, and $0.12$ better than the posted prices of the settled trades under FBA with an auction frequency of 5, 10, and 15 seconds. 

In summary, for the two trading targets during the sampled time window, FBA reduces welfare loss and saves transaction costs for users. \tian{Note that for the lack of more suitable datasets, we acquired the tick-level order book snapshots and transactions from an exchange that utilizes the CLOB processing model. If FBA were adopted, more transactions could have been settled since placed orders can be matched against each other, while the order books could have had different spreads.}

\subsection{An example with small parameters}\label{sec: example}

{While previous comparisons and empirical analysis are relatively thorough,
one thing they lack is reflecting how all different parameters in \Cref{thm:clob,thm: fba} interact.
To this end, we provide a small-parameter example as a demonstration.}

\paragraph{Welfare loss}
Without loss of generality, we consider a unit order for CLOB and a specific jump size $J$. 
Under CLOB, let $a = 1 + p^{\star} \frac{\mathsf{g}_{1} }{\mathsf{g}_{1} (r-2) + 1 } $ and the spread in \Cref{eq:spread} satisfies $\frac{s}{2} = \frac{(\player{pb} + a \player{pr}) J - (\player{i} + \player{pr}) \mathsf{g}_{1} \mathsf{F} }{\player{pb} + a \player{pr} + \player{i}} $. And the price impact from informed trader is $\Delta = J \frac{\player{pr}}{\player{pr} + \player{i}}$. Then the markup can be computed as $M_{C} = J \frac{\player{i} \player{pb} + (a - 1) \player{i} \player{pr} }{(\player{i} + a \player{pr} + \player{pb})(\player{pr} + \player{i})} - \frac{(\player{i} + \player{pr}) \mathsf{g}_{1}\mathsf{F} }{\player{pb} + a \player{pr} + \player{i}}$ 
\FULL{\small
\[ M_{C} = J \frac{\player{i} \player{pb} + (a - 1) \player{i} \player{pr} }{(\player{i} + a \player{pr} + \player{pb})(\player{pr} + \player{i})} - \frac{(\player{i} + \player{pr}) \mathsf{g}_{1}\mathsf{F} }{\player{pb} + a \player{pr} + \player{i}} \] 
}
and the expected markup is $J \frac{\player{i} \player{pb} + (a - 1) \player{i} \player{pr} }{\player{i} + a \player{pr} + \player{pb}} - \frac{(\player{i} + \player{pr})^2 \mathsf{g}_{1}\mathsf{F} }{\player{pb} + a \player{pr} + \player{i}}$. This increases with the jump size $J$ and decreases with priority fee $\mathsf{F} $. 
For simplicity, we can measure $\mathsf{F} $ relative to the jump, i.e., let $\mathsf{F} =x J$ for some $x<1$. 

Under FBA, we have that $\Delta_k = J \frac{\player{pr}}{\player{pr} + \player{i}}$. 
Consider $Q=2$. The expected markups become $\frac{2}{I} J \frac{\player{pr}}{\player{pr} + \player{i}} (q_1 (\alpha_1 +\alpha_1 \alpha_2) + 2 q_2 \alpha_2 ) $. 
For simplicity and without loss of generality, we let $q_3=q_2=1/16, q_1=1/8$, FBA is better if 
$\frac{\player{i} \player{pb} + (a-1) \player{i} \player{pr} - (\player{i} +\player{pr})^2 \mathsf{g}_{1} x}{\player{i} +a\player{pr}+\player{pb}} > \frac{\player{pr}}{I(\player{pr}+\player{i})} $, i.e., 

\small
\begin{multline*}
    \player{pb} > \frac{ \player{pr} (\player{i} + a \player{pr})}{I \player{i} (\player{pr}+\player{i})- \player{pr}} - \\
    \frac{ I (\player{i} + \player{pr}) [ (a-1) \player{i} \player{pr} - (\player{i} +\player{pr})^2 \mathsf{g}_{1} x]}{I \player{i} (\player{pr}+\player{i})- \player{pr}}
\end{multline*}
\normalsize

\begin{figure}[!htbp]
    \centering
    \begin{subfigure}[b]{0.3\textwidth}
        \centering
        \includegraphics[width=0.7\linewidth]{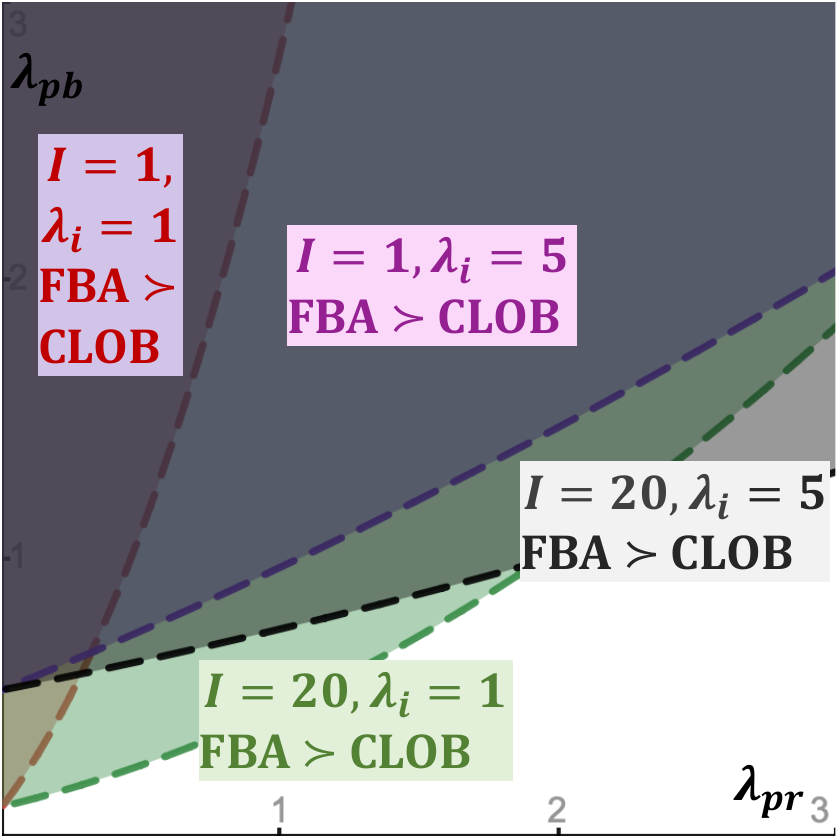}
        \caption{Example regions where FBA has less welfare loss when $\mathsf{F} =0.15J, p^*=0.8, r=35$, truncated at $\player{pr}=3, \player{pb}=3 $.
        }\label{fig: markups regions}
    \end{subfigure}
    \hfill
    \begin{subfigure}[b]{0.3\textwidth}
        \centering
        \includegraphics[width=0.7\linewidth]{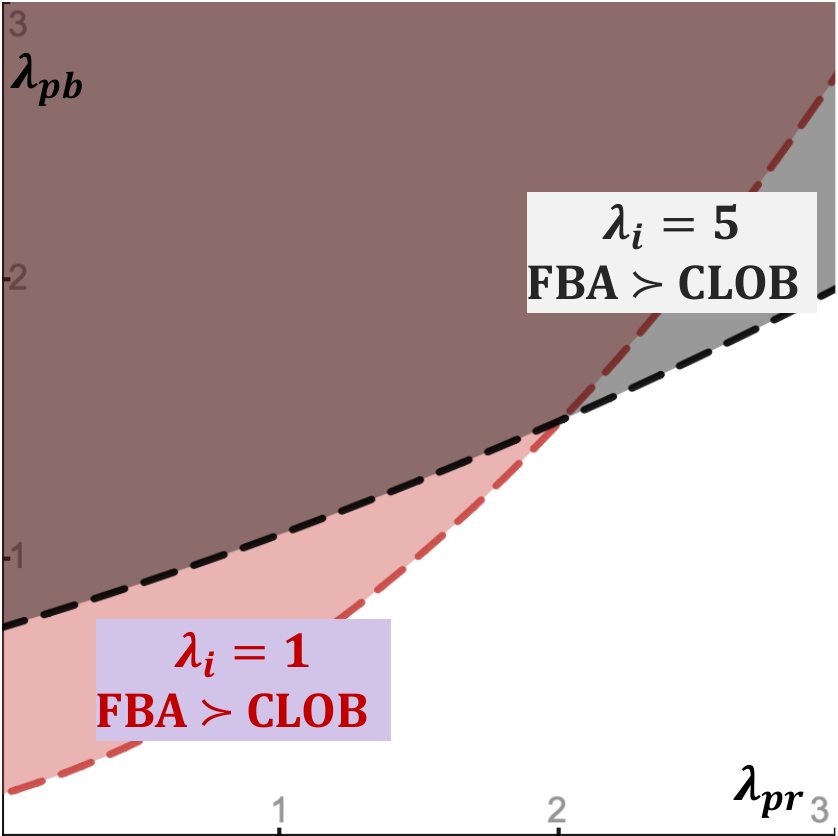}
        \caption{Example regions where FBA has smaller bid-ask spreads under the same parameters as \Cref{fig: markups regions} and additionally, $q_0=0.5, Q=100$. 
    }\label{fig: spreads regions}
    \end{subfigure}
    \caption{Examples with small parameters. The dashed lines represent that given certain $I, \player{i}$, the region above the line means FBA performs better than CLOB, and vise versa.}
\end{figure}

Note that $(q_1 + q_2 + q_3)$ is smaller than $1$ because the excess demand follows the Skellam distribution \tian{(since investor and trader arrivals are Poison processes, and they buy or sell equally likely)}. Intuitively, users' transactions can be executed against each other and the LPs only need to satisfy excessive demands. 
We depict the regions where FBA induces less welfare loss for common investors and traders in \Cref{fig: markups regions} in \FULL{(The dashed lines represent that given certain $I, \player{i}$, the region above the line means FBA performs better than CLOB, and vise versa.)}. We intentionally let $x$ or $\mathsf{F} $ be not too small, i.e., $x=0.15$, in the figure to make the regions where CLOB imposes fewer markups more visible\FULL{ in this example}. This means that if the jump $J=\$1$, the priority fee is $\$0.15$. When the priority fee $\mathsf{F} $ decreases, the region where FBA has fewer markups expands.

\paragraph{Spread}
In the same setting with constant jump size, under CLOB, the bid-ask spread is $s = \frac{(\player{pb} + a \player{pr}) J - (\player{i} + \player{pr}) \mathsf{g}_{1} \mathsf{F} }{\player{pb} + a \player{pr} + \player{i}}$. Under FBA, we denote the bid-ask spread as $s_{F}$ and $s_{F} = 2(\Delta + M_1)$. Consider $Q\geq 1$. Same as before, although $Z_I$ follows a Skellam distribution, for simplicity and without loss of generality, we consider $q_1=\frac{1}{8}$ and $q_2=\cdots = q_{Q+1}=\frac{1}{8Q}$. Then 
\small
\begin{align*}
 M_1 &= \sum_{u=1}^Q \Delta_u \prod_{v=1}^{u} \alpha_{v} = \Delta (\alpha_1 + \alpha_1 \alpha_2 + \ldots + \alpha_1 \cdots \alpha_{Q}) \\
 &= \Delta \frac{\frac{1}{8Q}}{\frac{1}{8} + \frac{1}{8Q}} (1 + \frac{1}{2} + \ldots + \frac{1}{2^{Q-1}}) = \Delta \frac{2}{Q+1} (1- \frac{1}{2^Q})
\end{align*}
\normalsize
Let a transaction be fulfilled with transactions submitted by other common investors or traders with probability $q_0 = 1 - 2 \sum_{j=1}^{Q+1} q_j$. 
FBA provides a smaller bid-ask spread in expectation when $(1- q_0) s > s_F$. This gives 
\small
\[
    2 (1 - q_0) \frac{\player{pr}}{\player{pr}+ \player{i}} (1 + \frac{2}{Q+1} (1- \frac{1}{2^Q})) < 
    \frac{\player{pb} + a \player{pr} - (\player{i} + \player{pr}) \mathsf{g}_{1} x }{\player{pb} + a \player{pr} + \player{i}}
\]
\normalsize

We depict the regions where FBA has a smaller expected bid-ask spread in \Cref{fig: spreads regions}. 
When there is no private information, the spread under FBA is constant while the spread under CLOB increases with $\player{pb}$. 
We can also consider general parameterization for $Z_I$'s distribution. When $\sum_{u=1}^Q \prod_{v=1}^{u} \alpha_{v}$ decreases, the region where FBA has smaller spreads expands. 
In summary, when public information is released more often or common investors arrive more often compared with privately informed traders, FBA yields less welfare loss than CLOB.

\section{Related work}\label{sec:litrature}
\paragraph{Batch auction-based DEX}
Penumbra~\cite{penumbra} adopts FBA as the processing model. CoW Swap~\cite{cow} combines batch auction with trade intents where users specify the assets and the amounts to trade. Users' intents are then settled via direct matching among intents for the same tokens and opposite sides.

\paragraph{Manipulation-aware DEX design} 
We mention mechanisms that update transaction \textit{execution} rules to mitigate order manipulation attacks in DEX because they are closely related to this work. For general order manipulation mitigation techniques that update \textit{ordering} algorithms, we refer the readers to comprehensive surveys like~\cite{heimbach2022sok}.
Fair-TraDEX~\cite{mcmenamin2022fairtradex} combines FBA with commit-then-reveal style transaction masking where users commit to transactions and later reveal the contents upon finalization. Each user needs to make deposits in an escrow service so that they are incentivized to post correct commitments. 
Similarly, Injective~\cite{injective} and Penumbra~\cite{penumbra} utilize FBA for executing transactions on proof-of-stake blockchains, with Penumbra being a private chain. It additionally hides transactions' trading amounts. 
Masking transaction limits may not eliminate all manipulation opportunities. 
In P2DEX~\cite{baum2021p2dex}, servers run \FULL{a }secure multi-party computation (MPC) \FULL{protocol }to match orders. This introduces computation overhead and latencies and adds constraints inherent to the adopted MPC\FULL{ machinery}. 
SPEEDEX~\cite{ramseyer2021speedex} approximates the clearing price in a block in general equilibrium where demand meets supply given a set of static orders. 
The equilibrium does not \FULL{aim to }capture participants' strategies, acquired information, or sequential moves. 
Xavier et al.~\cite{xavier2023credible} alternatively consider a greedy sequencing rule in two-token constant product \FULL{automatic market maker}\CONF{AMM}. 
Since this sequencing rule ensures good properties only inside a block, sequencers can push submitted transactions to future blocks. 

\paragraph{Comparison of CLOB and FBA in centralized exchanges}
It has been shown that compared with CLOB, FBA leads to lower transaction costs~\cite{aldrich2020experiments}, decreases adverse selection and spreads~\cite{menkveld2017need,ricco2020frequent}, achieves an optimal trade-off between liquidity and price discovery~\cite{baldauf2020high}, and increases market quality~\cite{madhavan1992trading,economides2001electronic,kandel2012effect}. 
However, the severity of the inefficiency of liquidity provision under FBA can exceed the inefficiency from latency arbitrage under CLOB~\cite{eibelshauser2022frequent}. This liquidity provision inefficiency of FBA originates from bid shading\footnote{(Non-unit-demand) bidders' tendency to bid less than true valuations for later units.} in \CONF{UPDA}\FULL{uniform price double auctions}: every equilibrium in multi-unit uniform price auctions is inefficient due to bid shading~\cite{ausubel2002demand}. 
\section*{Acknowledgement}
This work was supported in part by the National Science Foundation (NSF) under grant CNS1846316 and a Purdue Research Foundation (PRF) research grant.

\bibliographystyle{ieeetr}
\bibliography{all}

\begin{thebibliography}{10}

\bibitem{sui}
M.~Labs, ``sui.'' \url{https://github.com/MystenLabs/sui}, 2023.

\bibitem{penumbra}
Penumbra, ``Penumbra zswap.''
\newblock \url{https://protocol.penumbra.zone}.

\bibitem{uniswaporderbook}
U.~Labs, ``Uniswap limit order book.''
  \url{https://blog.uniswap.org/limit-orders}, 2024.

\bibitem{cow}
CoW, ``Cow protocol.''
\newblock \url{https://docs.cow.fi/cow-protocol}.

\bibitem{adams2021uniswap}
H.~Adams, N.~Zinsmeister, M.~Salem, R.~Keefer, and D.~Robinson, ``Uniswap v3
  core,'' {\em Tech. rep., Uniswap, Tech. Rep.}, 2021.

\bibitem{farmer2012review}
D.~Farmer and S.~Skouras, ``Review of the benefits of a continuous market vs.
  randomised stop auctions and of alternative priority rules (policy options 7
  and 12),'' {\em Manuscript, Foresight. Government Office for Science}, 2012.

\bibitem{budish2015high}
E.~Budish, P.~Cramton, and J.~Shim, ``The high-frequency trading arms race:
  Frequent batch auctions as a market design response,'' {\em The Quarterly
  Journal of Economics}, vol.~130, no.~4, pp.~1547--1621, 2015.

\bibitem{twse}
T.~S.~E. Corporation, ``Taiwan stock exchange.''
\newblock \url{https://www.twse.com.tw/en/}.

\bibitem{budish2014implementation}
E.~Budish, P.~Cramton, and J.~Shim, ``Implementation details for frequent batch
  auctions: Slowing down markets to the blink of an eye,'' {\em American
  Economic Review}, vol.~104, no.~5, pp.~418--424, 2014.

\bibitem{glosten1985bid}
L.~R. Glosten and P.~R. Milgrom, ``Bid, ask and transaction prices in a
  specialist market with heterogeneously informed traders,'' {\em Journal of
  financial economics}, vol.~14, no.~1, pp.~71--100, 1985.

\bibitem{ethfee}
ethereum.org, ``Priority fees in ethereum.''
\newblock \url{https://ethereum.org/en/developers/docs/gas/#priority-fee}.

\bibitem{dydx}
A.~Juliano, ``dydx.'' \url{https://github.com/dydxprotocol}, 2023.

\bibitem{cachin2021quick}
C.~Cachin, J.~Mi{\'c}i{\'c}, N.~Steinhauer, and L.~Zanolini, ``Quick order
  fairness,'' in {\em FC}, pp.~316--333, 2022.

\bibitem{eibelshauser2022frequent}
S.~Eibelsh{\"a}user and F.~Smetak, ``Frequent batch auctions and informed
  trading,'' 2022.

\bibitem{maskin2001markov}
E.~Maskin and J.~Tirole, ``Markov perfect equilibrium: I. observable actions,''
  {\em Journal of Economic Theory}, vol.~100, no.~2, pp.~191--219, 2001.

\bibitem{budish2019theory}
E.~Budish, R.~S. Lee, and J.~J. Shim, ``A theory of stock exchange competition
  and innovation: Will the market fix the market?,'' tech. rep., National
  Bureau of Economic Research, 2019.

\bibitem{tardis}
Tardis.dev, ``Tardis.dev node.''
  \url{https://github.com/tardis-dev/tardis-node}, 2023.

\bibitem{simulationsrc}
Anonymized, ``Simulation source code.''
\newblock
  \url{https://drive.google.com/drive/folders/1XtnstLf5oOY8KTp_YMOwpNbHgs01JpGC?usp=share_link}.

\bibitem{indriawan2020effects}
I.~Indriawan, R.~Pascual, and A.~Shkilko, ``On the effects of continuous
  trading,'' {\em Available at SSRN 3707154}, 2020.

\bibitem{heimbach2022sok}
L.~Heimbach and R.~Wattenhofer, ``Sok: Preventing transaction reordering
  manipulations in decentralized finance,'' in {\em Proceedings of the 4th ACM
  Conference on Advances in Financial Technologies}, pp.~47--60, 2022.

\bibitem{mcmenamin2022fairtradex}
C.~McMenamin, V.~Daza, M.~Fitzi, and P.~O'Donoghue, ``Fairtradex: A
  decentralised exchange preventing value extraction,'' in {\em ACM CCS DeFi
  Workshop'22}, pp.~39--46, 2022.

\bibitem{injective}
I.~Labs, ``Injective.''
\newblock \url{https://injective.com/about}.

\bibitem{baum2021p2dex}
C.~Baum, B.~David, and T.~K. Frederiksen, ``P2dex: privacy-preserving
  decentralized cryptocurrency exchange,'' in {\em ACNS}, pp.~163--194, 2021.

\bibitem{ramseyer2021speedex}
G.~Ramseyer, A.~Goel, and D.~Mazi{\`e}res, ``Speedex: A scalable,
  parallelizable, and economically efficient digital exchange,'' {\em arXiv
  preprint arXiv:2111.02719}, 2021.

\bibitem{xavier2023credible}
M.~V. Xavier~Ferreira and D.~C. Parkes, ``Credible decentralized exchange
  design via verifiable sequencing rules,'' in {\em Proceedings of the 55th
  Annual ACM Symposium on Theory of Computing}, pp.~723--736, 2023.

\bibitem{aldrich2020experiments}
E.~M. Aldrich and K.~L{\'o}pez~Vargas, ``Experiments in high-frequency trading:
  comparing two market institutions,'' {\em Experimental Economics}, vol.~23,
  no.~2, pp.~322--352, 2020.

\bibitem{menkveld2017need}
A.~J. Menkveld and M.~A. Zoican, ``Need for speed? exchange latency and
  liquidity,'' {\em The Review of Financial Studies}, vol.~30, no.~4,
  pp.~1188--1228, 2017.

\bibitem{ricco2020frequent}
R.~Ricc{\`o} and K.~Wang, ``Frequent batch auctions vs. continuous trading:
  Evidence from taiwan,'' {\em Continuous Trading: Evidence from Taiwan
  (November 19, 2020)}, 2020.

\bibitem{baldauf2020high}
M.~Baldauf and J.~Mollner, ``High-frequency trading and market performance,''
  {\em The Journal of Finance}, vol.~75, no.~3, pp.~1495--1526, 2020.

\bibitem{madhavan1992trading}
A.~Madhavan, ``Trading mechanisms in securities markets,'' {\em the Journal of
  Finance}, vol.~47, no.~2, pp.~607--641, 1992.

\bibitem{economides2001electronic}
N.~Economides and R.~A. Schwartz, ``Electronic call market trading,'' in {\em
  The Electronic Call Auction: Market Mechanism and Trading}, pp.~87--99,
  Springer, 2001.

\bibitem{kandel2012effect}
E.~Kandel, B.~Rindi, and L.~Bosetti, ``The effect of a closing call auction on
  market quality and trading strategies,'' {\em Journal of Financial
  Intermediation}, vol.~21, no.~1, pp.~23--49, 2012.

\bibitem{ausubel2002demand}
L.~M. Ausubel and P.~Cramton, ``Demand reduction and inefficiency in multi-unit
  auctions,'' 2002.

\end{thebibliography}

\appendix
\section{Bid-ask spread comparison figures}\label{app: liquidity comparison}
\begin{figure*}[!htbp]
    \centering
    \begin{subfigure}[b]{0.48\textwidth}
        \centering
        \includegraphics[width=\textwidth]{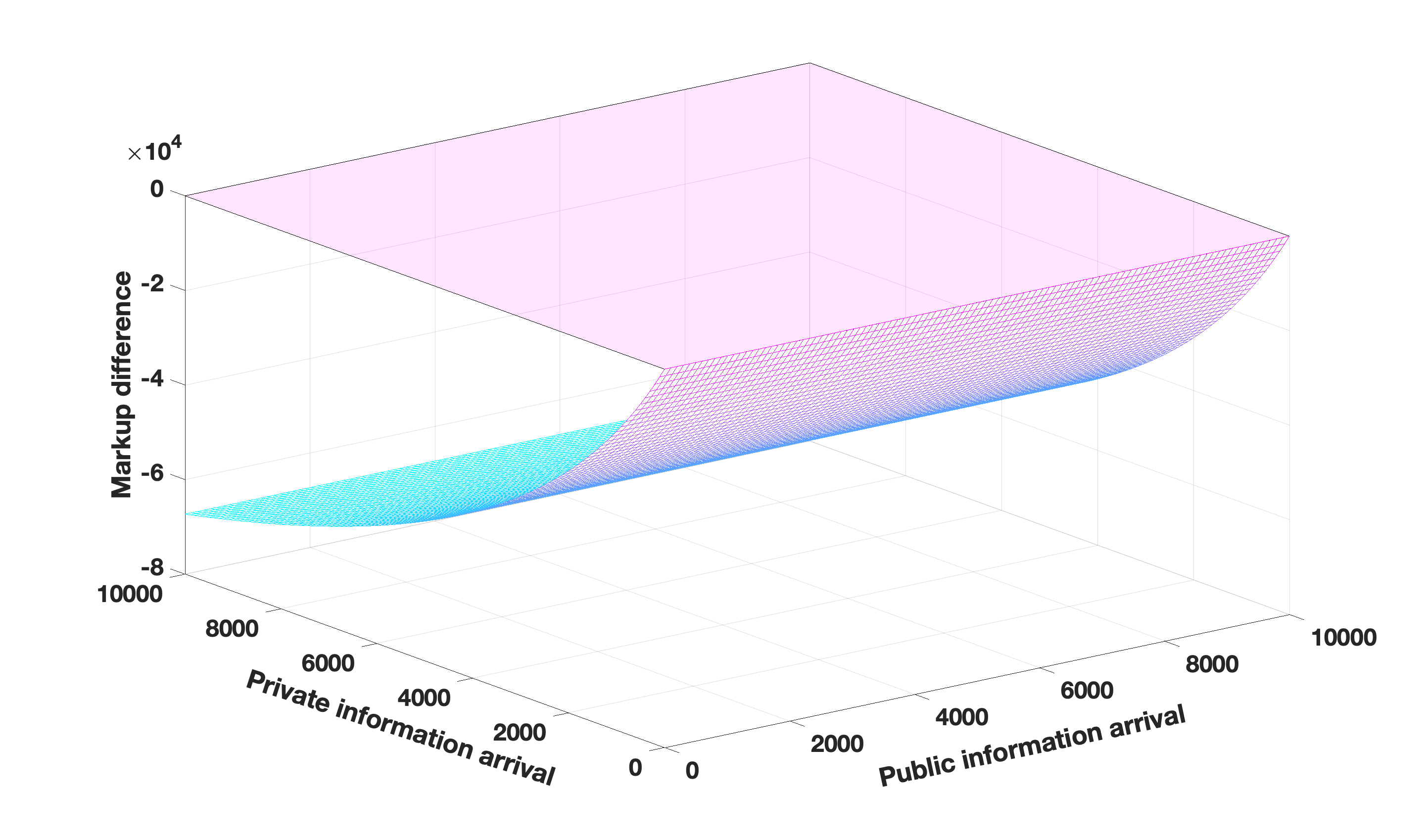}
        \caption{Bad case for both CLOB and FBA.}
    \end{subfigure}
    \hfill
    \begin{subfigure}[b]{0.48\textwidth}
        \centering
        \includegraphics[width=\textwidth]{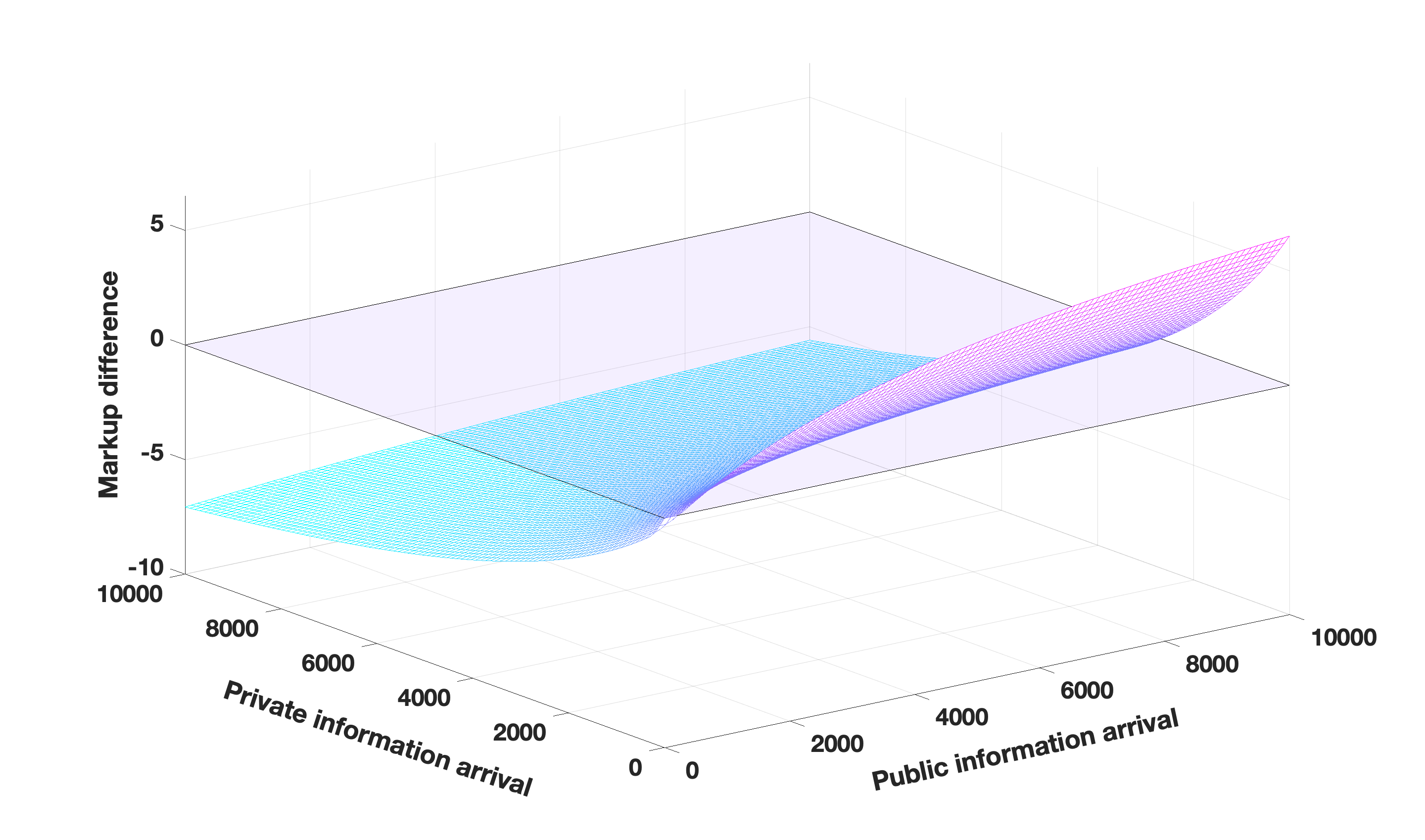}
        \caption{Bad case for CLOB and good case for FBA.}
    \end{subfigure}
    \begin{subfigure}[b]{0.48\textwidth}
        \centering
        \includegraphics[width=\textwidth]{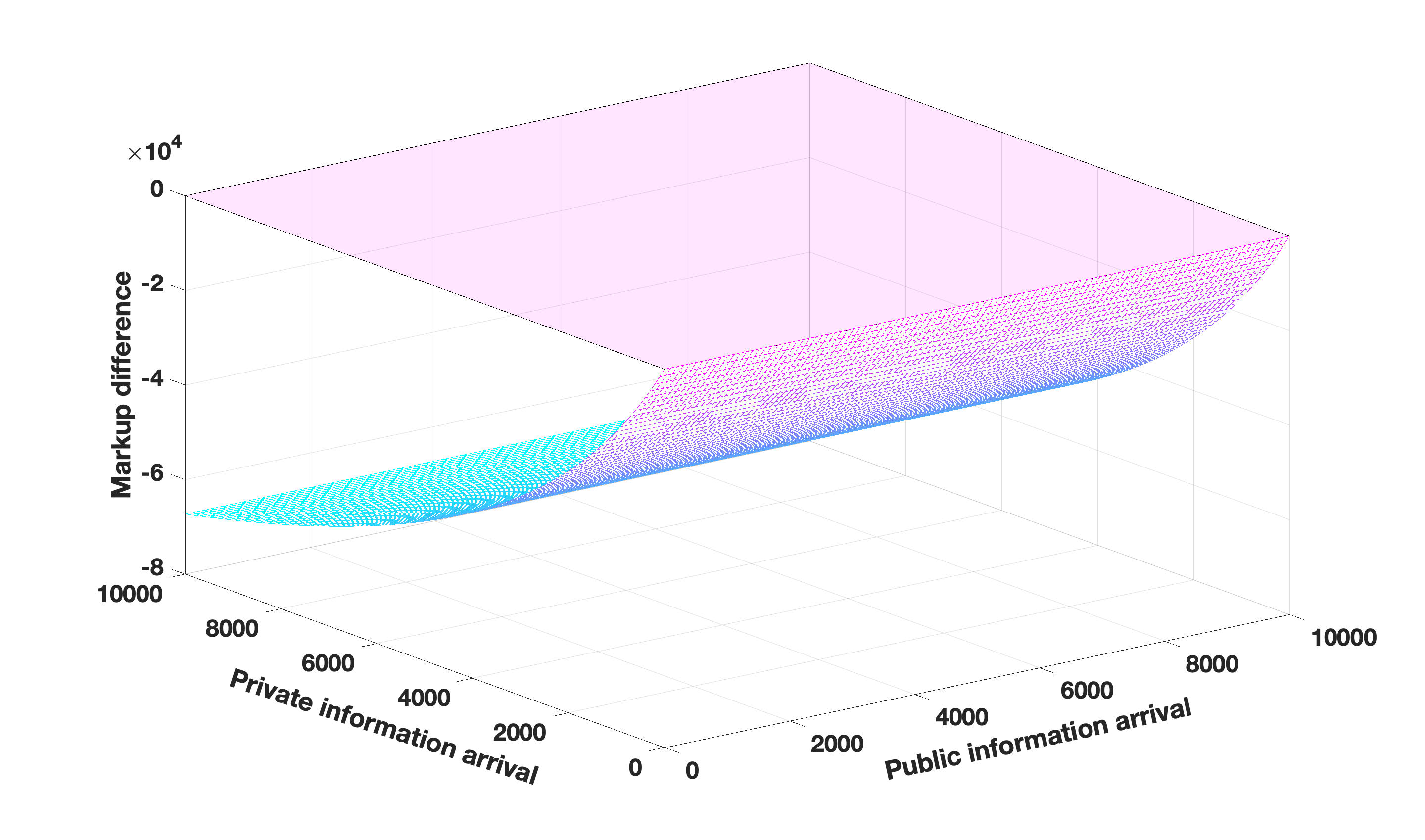}
        \caption{Good case for CLOB and bad case for FBA.}
    \end{subfigure}
    \hfill
    \begin{subfigure}[b]{0.48\textwidth}
        \centering
        \includegraphics[width=\textwidth]{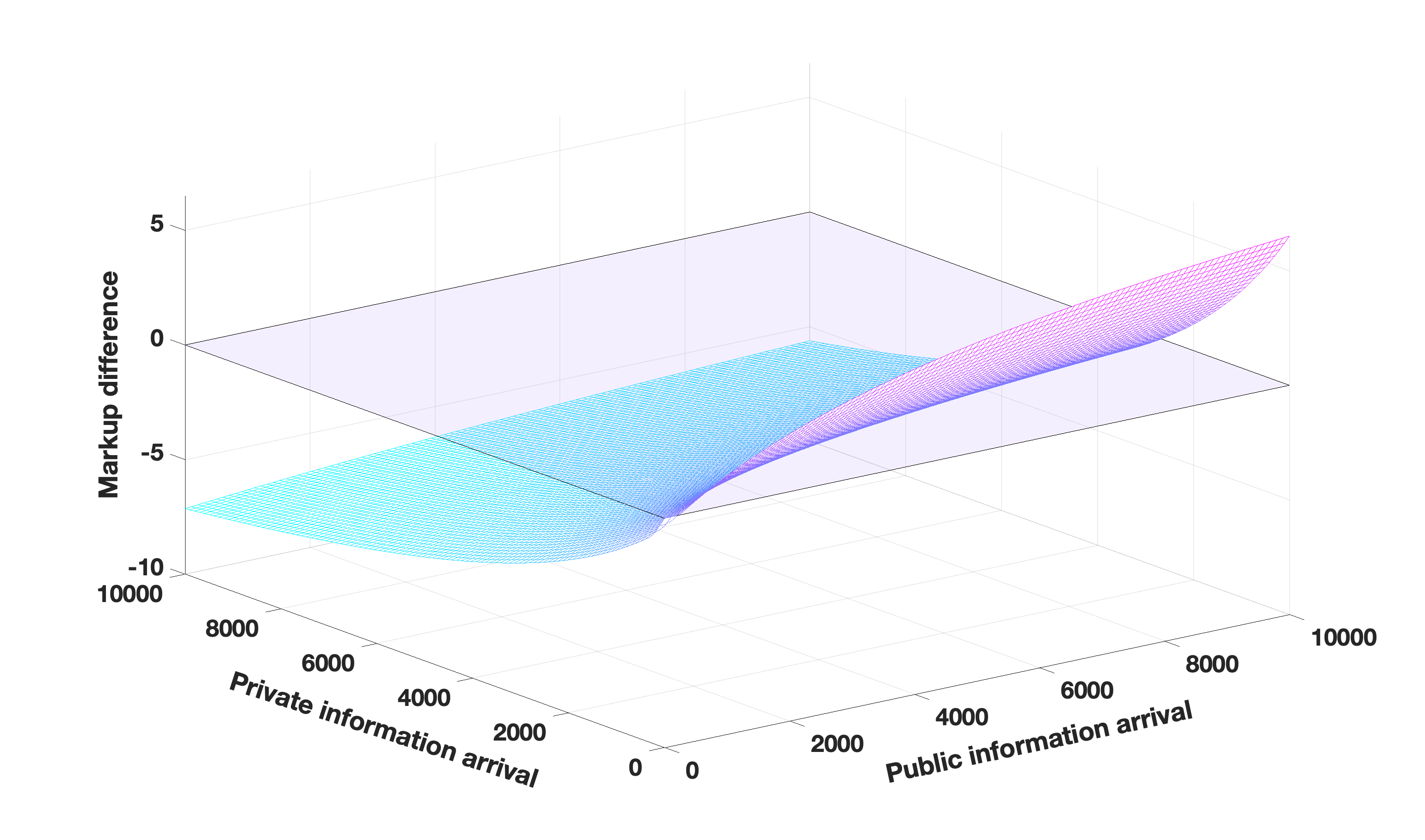}
        \caption{Good case for both CLOB and FBA.}
    \end{subfigure}
    \caption{The spread difference between CLOB and FBA with respect to public and private information arrivals. The investor arrival rate $\player{i}$ is set to be $5000$. Increasing (decreasing) $\player{i}$ pushes the surface up (down).
    }\label{fig: spread pbpr}

    \begin{subfigure}[b]{0.48\textwidth}
        \centering
        \includegraphics[width=\textwidth]{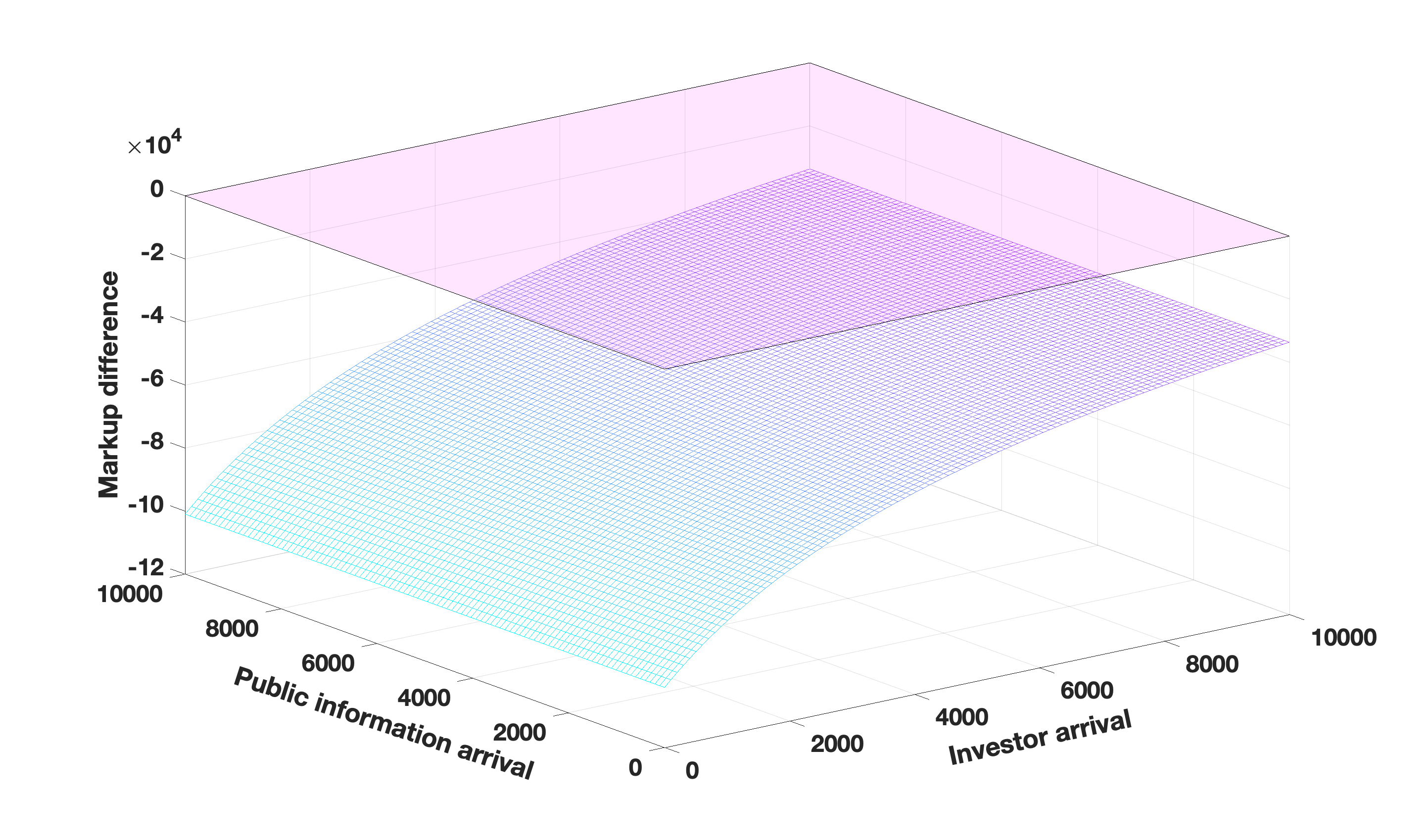}
        \caption{Bad case for both CLOB and FBA.}
    \end{subfigure}
    \hfill
    \begin{subfigure}[b]{0.48\textwidth}
        \centering
        \includegraphics[width=\textwidth]{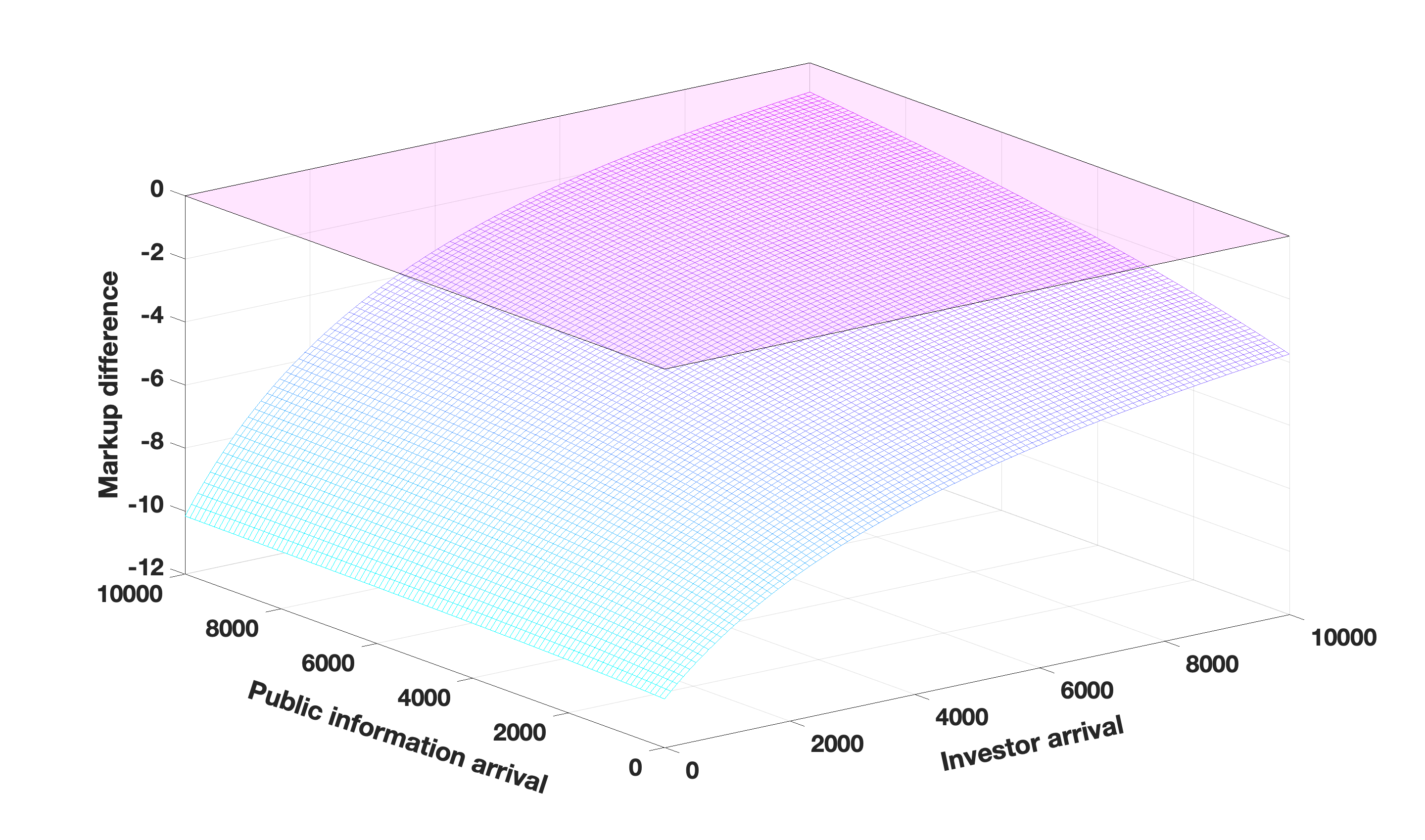}
        \caption{Bad case for CLOB and good case for FBA.}
    \end{subfigure}
    \begin{subfigure}[b]{0.48\textwidth}
        \centering
        \includegraphics[width=\textwidth]{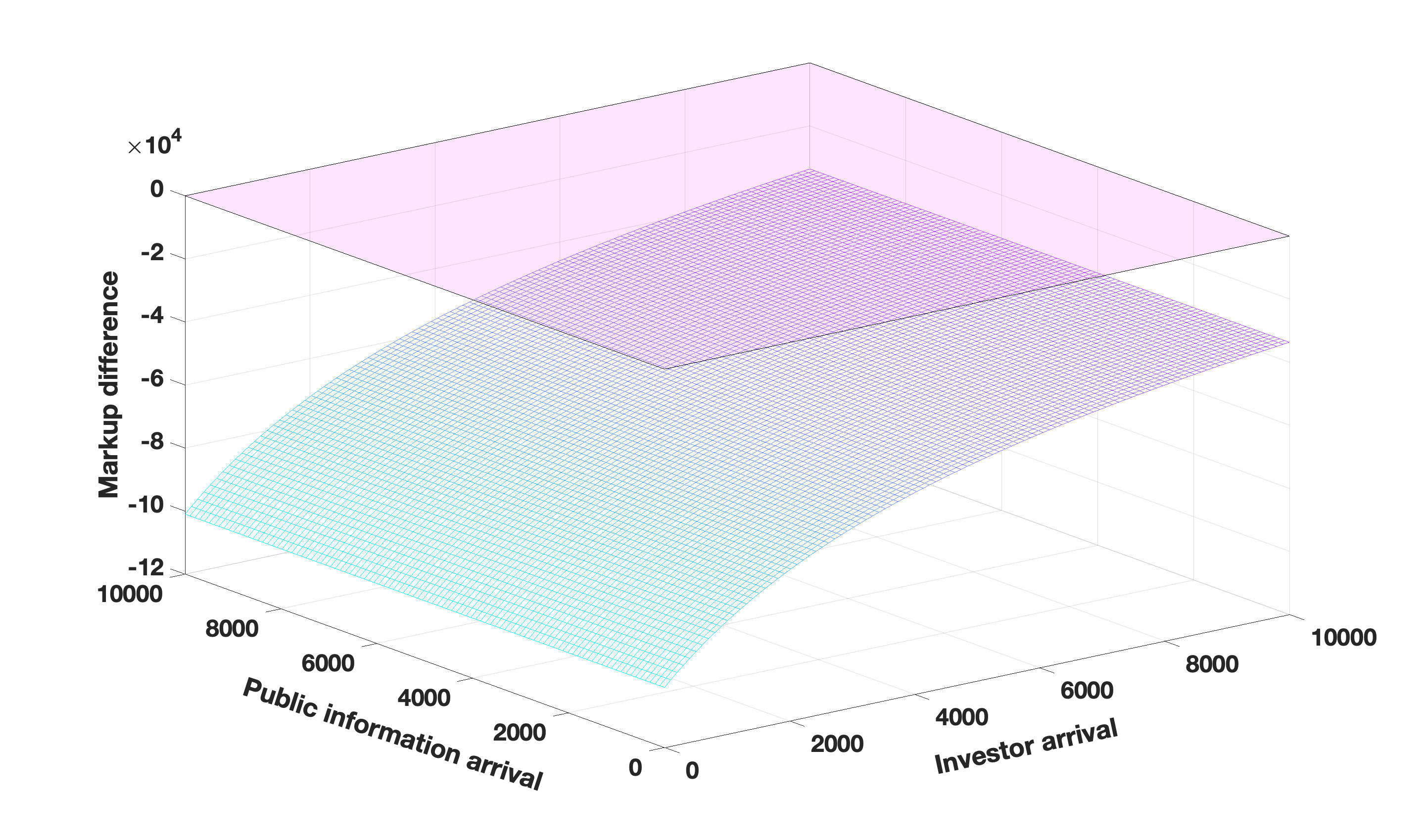}
        \caption{Good case for CLOB and bad case for FBA.}
    \end{subfigure}
    \hfill
    \begin{subfigure}[b]{0.48\textwidth}
        \centering
        \includegraphics[width=\textwidth]{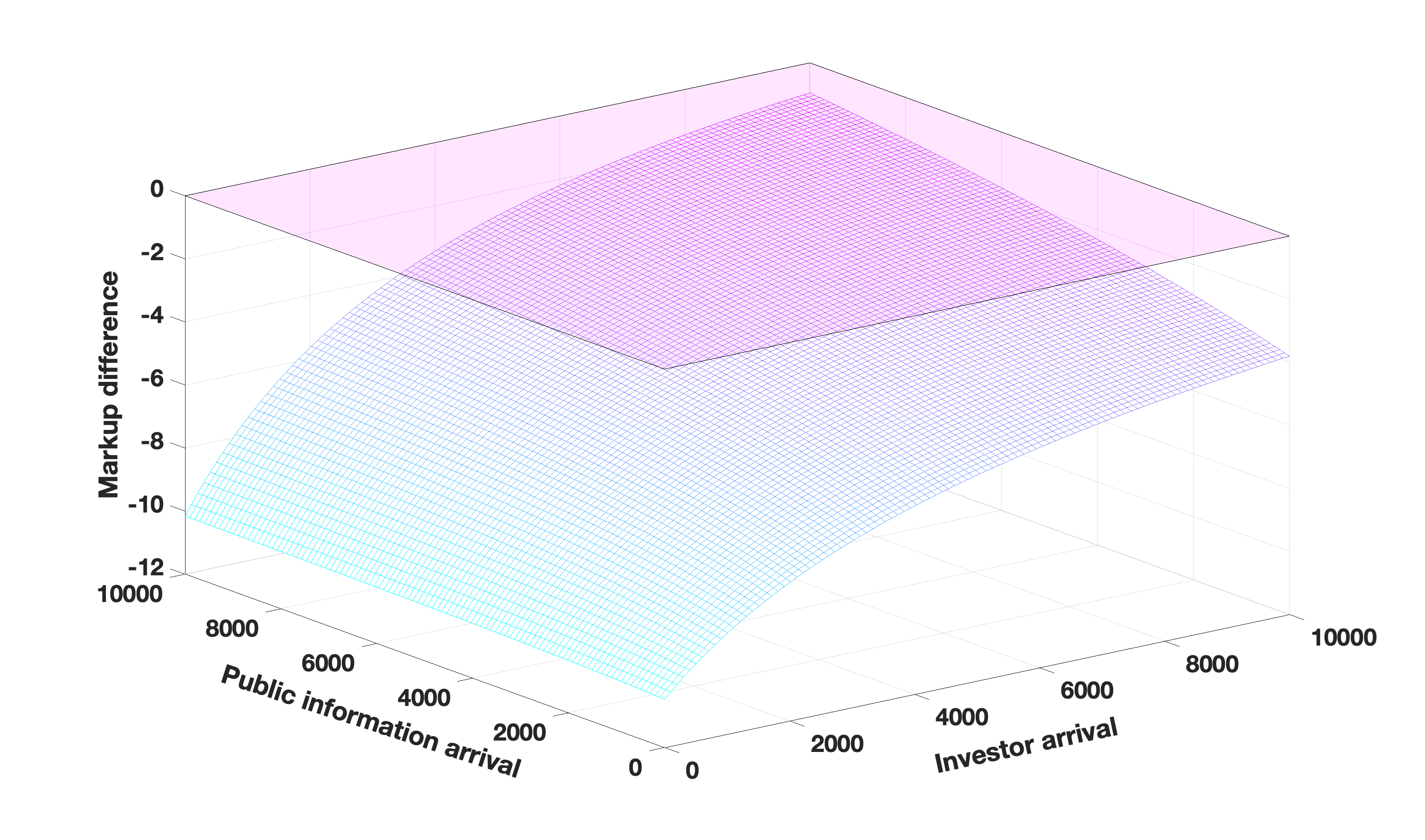}
        \caption{Good case for both CLOB and FBA.}
    \end{subfigure}
    \caption{The spread difference between CLOB and FBA with respect to public and private information arrivals. The private information arrival rate $\player{pr}$ is set to be $5000$. Increasing (decreasing) $\player{pb}$ pushes the surface down (up).}\label{fig: spread ipb}
\end{figure*}

\begin{figure*}[!htbp]
    \centering
    \begin{subfigure}[b]{0.48\textwidth}
        \centering
        \includegraphics[width=\textwidth]{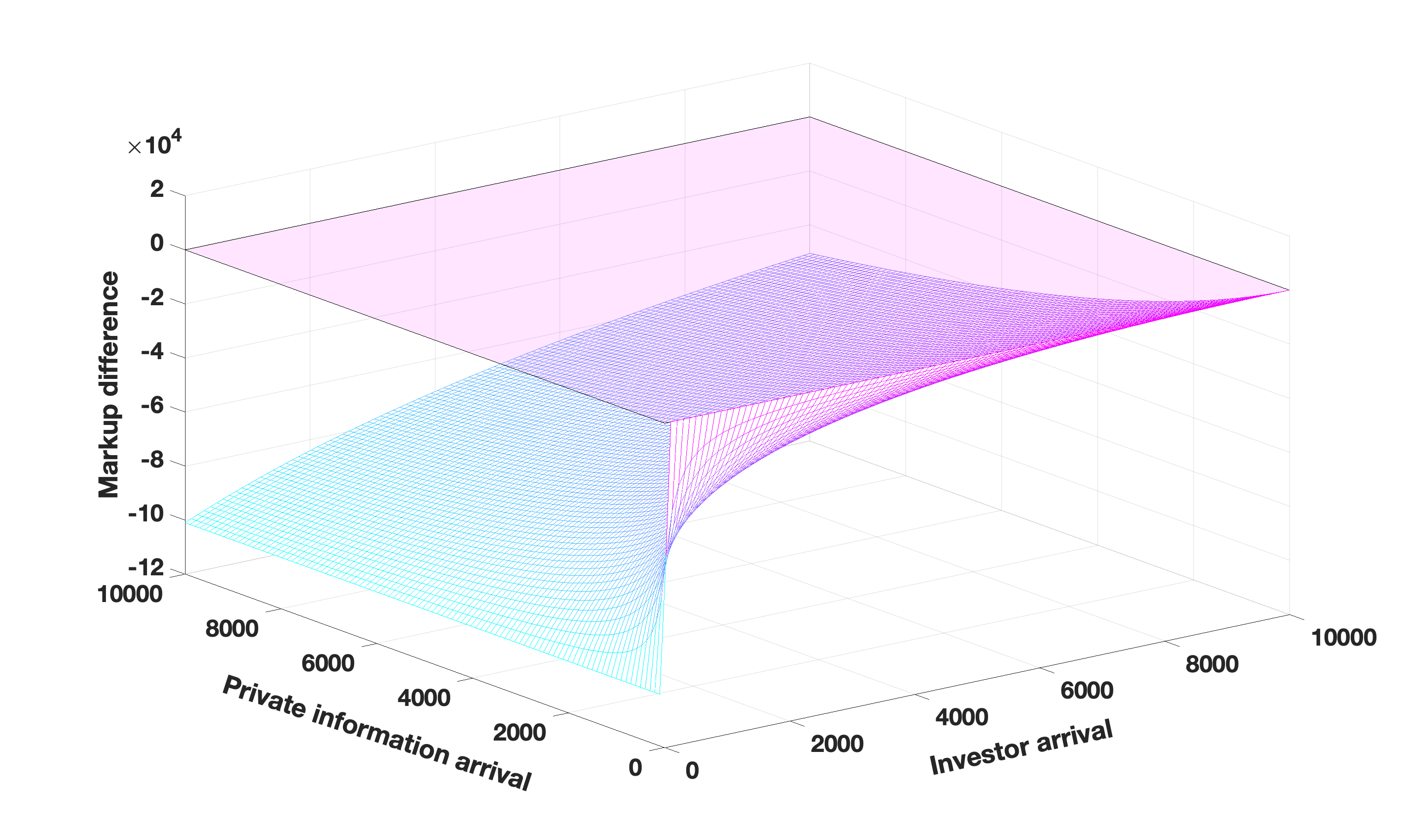}
        \caption{Bad case for both CLOB and FBA.}
    \end{subfigure}
    \hfill
    \begin{subfigure}[b]{0.48\textwidth}
        \centering
        \includegraphics[width=\textwidth]{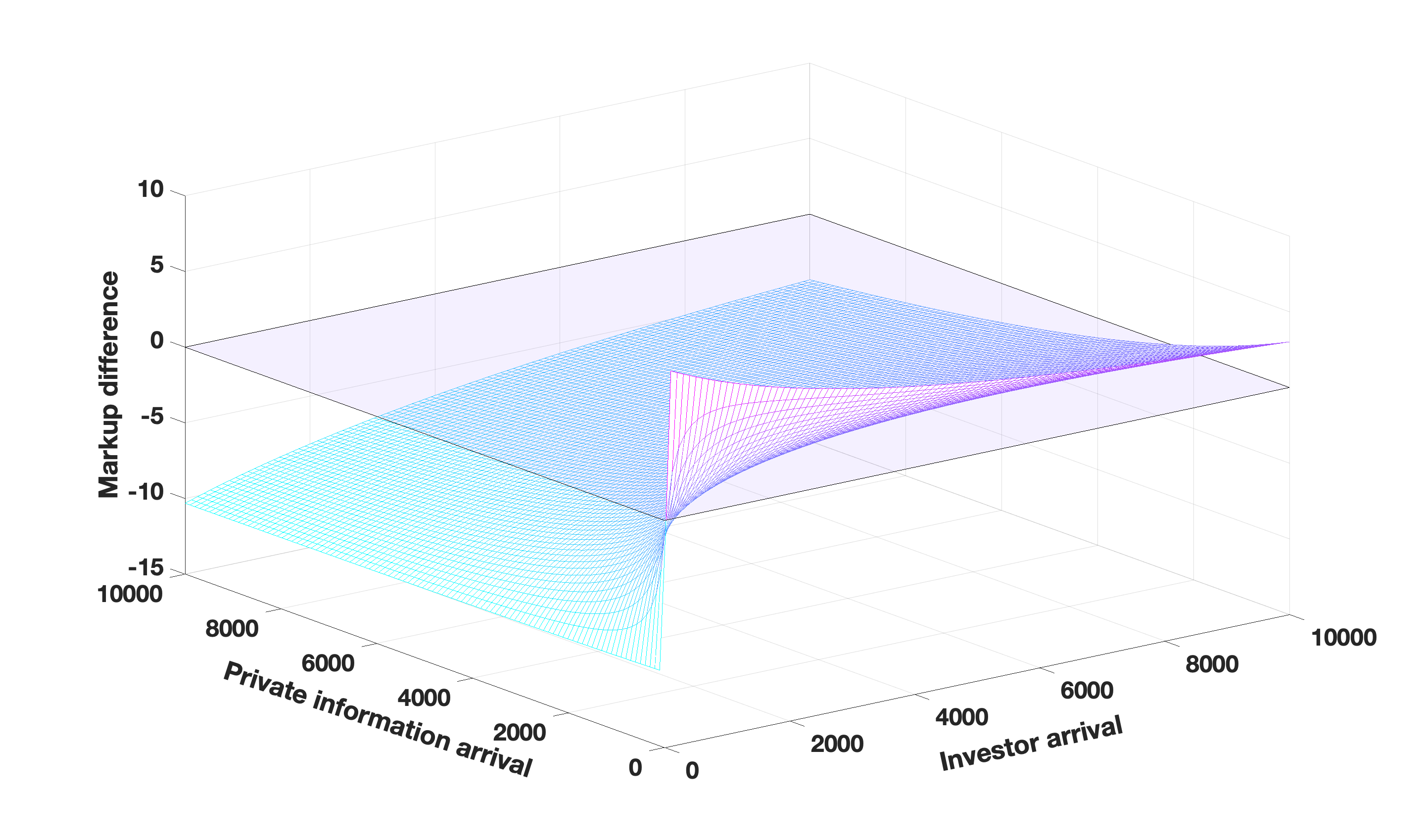}
        \caption{Bad case for CLOB and good case for FBA.}
    \end{subfigure}
    \begin{subfigure}[b]{0.48\textwidth}
        \centering
        \includegraphics[width=\textwidth]{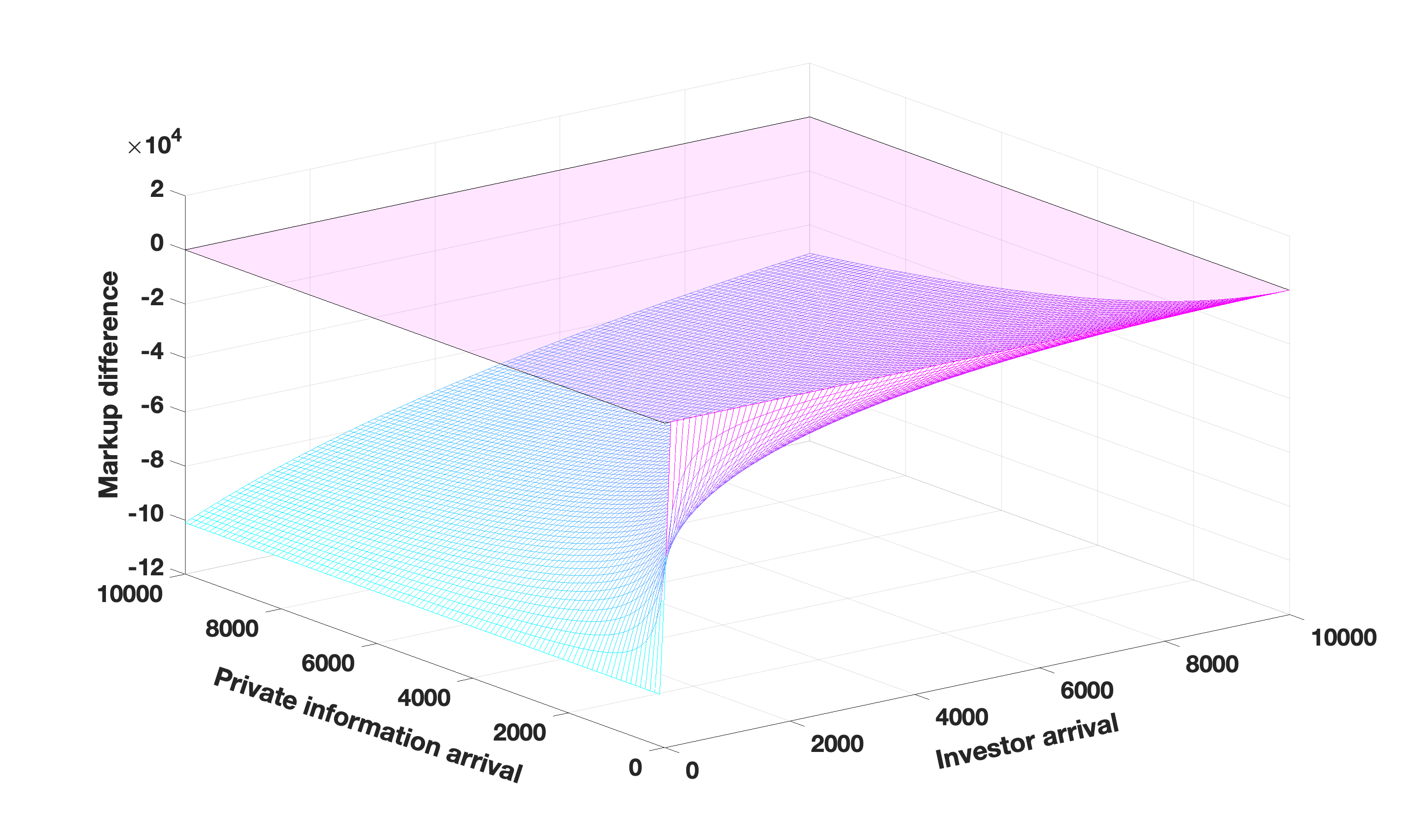}
        \caption{Good case for CLOB and bad case for FBA.}
    \end{subfigure}
    \hfill
    \begin{subfigure}[b]{0.48\textwidth}
        \centering
        \includegraphics[width=\textwidth]{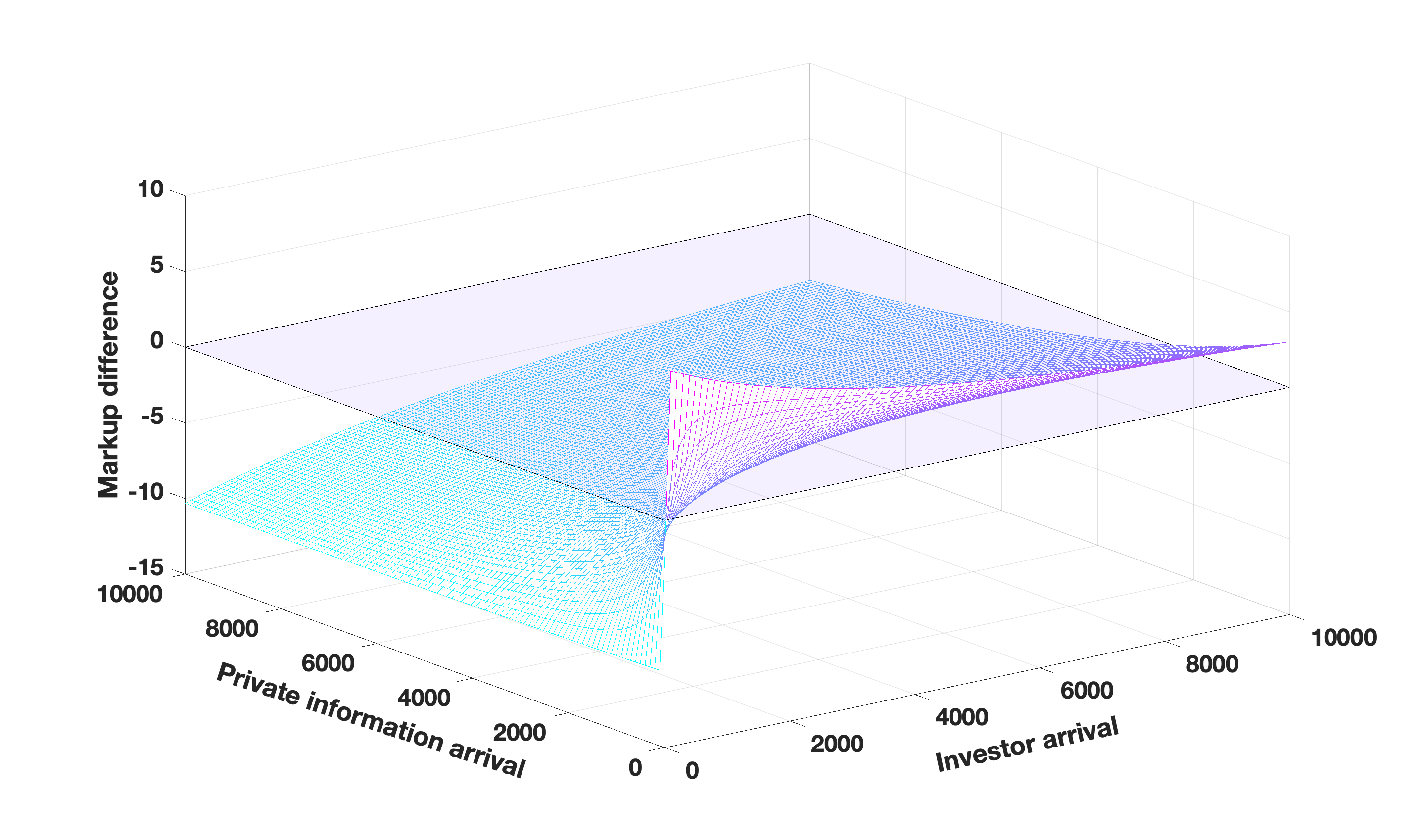}
        \caption{Good case for both CLOB and FBA.}
    \end{subfigure}
    \caption{The spread difference between CLOB and FBA with respect to public and private information arrivals. The public information arrival rate $\player{pb}$ is set to be $5000$. Increasing (decreasing) $\player{pb}$ pushes the surface up (down).
    }\label{fig: spread ipr}
\end{figure*}

We include the figures (\Cref{fig: spread pbpr,fig: spread ipb,fig: spread ipr}) depicting how $\player{i}, \player{pb}, \player{pr}$ affect the spreads in FBA and CLOB, as mentioned in \cref{sec: liquidity comparison}.

\end{document}